\documentclass{LMCS}

\def\doi{8 (1:06) 2012}
\lmcsheading%
{\doi}
{1--23}
{}
{}
{Nov.~21, 2009}
{Feb.~16, 2012}
{}

\usepackage{enumerate}
\usepackage{hyperref}
\usepackage{amsmath}
\usepackage{amssymb}
\usepackage{stmaryrd}
\usepackage{color}
\usepackage{xspace}
\usepackage{graphics}
\usepackage{epsfig}
\usepackage{xy} \xyoption{all} \CompileMatrices
\usepackage{proof}

\def\eg{{\em e.g.}}
\def\ie{{\em i.e.}}

\def\prem#1{\PP(#1)}

\def\ione{(\textbf{i1})}

\def\cczero{(\textbf{cc0})}
\def\ccone{(\textbf{cc1})}
\def\cctwo{(\textbf{cc2})}

\newcommand{\eq}[1]{\llbracket#1\rrbracket}

\newcommand{\G}{G}

\newcommand{\pr}[1]{\mathsf{P}_{#1}}

\newcommand{\thmref}[1]{The\-o\-rem~\ref{#1}}
\newcommand{\secref}[1]{\S\ref{#1}}
\newcommand{\lemref}[1]{Lemma~\ref{#1}}

\newcommand{\figref}[1]{Fig\-ure~\ref{#1}}
\newcommand{\exref}[1]{Ex\-ample~\ref{#1}}

\def\Gamma{\varGamma}
\def\Delta{\varDelta}
\def\Phi{\varPhi}
\def\Psi{\varPsi}
\def\emptyset{\varnothing}
\def\Sigma{\varSigma}
\def\Xi{\varXi}

\def\ignore#1{}

\newcommand{\defemph}[1]{\textsc{#1}}

\newcommand{\EUF}{\ensuremath{\mathit{EUF}}\xspace}

\newcommand{\eqs}[1]{\stackrel{#1}{\leftrightarrow}}

\newcommand{\D}{\mathsf{D}}
\newcommand{\limplies}{\Rightarrow}

\def\AA{\mathcal A}
\newcommand{\BB}{\ensuremath{\mathcal B}\xspace}
\newcommand{\II}{\ensuremath{\mathcal I}\xspace}
\newcommand{\PP}{\ensuremath{\mathcal P}\xspace}

\newcommand{\TT}{\ensuremath{\mathcal T}\xspace}
\newcommand{\RR}{\ensuremath{\mathcal R}\xspace}
\newcommand{\FF}{\ensuremath{\mathcal F}\xspace}

\newcommand{\EE}{\ensuremath{\mathcal E}\xspace}

\newcommand{\tentails}{\models_\TT}

\def\constname#1{\ensuremath{\mathsf{#1}}}
\newcommand{\true}{\constname{true}}
\newcommand{\false}{\constname{false}}

\newcommand{\ac}[1]{\textcolor{blue}{#1}}
\newcommand{\bc}[1]{\textcolor{red}{#1}}

\def\path#1#2{\overline{#1#2}}

\def\eqs#1{\llbracket{#1}\rrbracket}

\newcommand{\changed}[1]{#1}

\begin{document}

\title[ Ground interpolation 
for the theory of equality]{Ground interpolation for the theory of equality\rsuper*}

\author[A.~Fuchs]{Alexander Fuchs\rsuper a}	
\address{{\lsuper{a,e}}Department of Computer Science \\
The University of Iowa}	
\email{cesare-tinelli@uiowa.edu}  

\author[A.~Goel]{Amit Goel\rsuper b}	
\address{{\lsuper{b,c,d}}Strategic CAD Labs\\
Intel Corporation}	
\email{\{amit1.goel,jim.d.grundy,sava.krstic\}@intel.com}  

\author[J.~Gundy]{Jim Grundy\rsuper c}	
\address{\vskip-6 pt}	

\author[S.~Krsti\'c]{Sava Krsti\'c\rsuper d}	
\address{\vskip-6 pt}	

\author[C.~Tinelli]{Cesare Tinelli\rsuper e}	
\address{\vskip-6 pt}	
\thanks{{\lsuper e}Partially supported AFOSR Grant FA9550-09-1-0517.}	



\titlecomment{
{\lsuper*}Revised and extended version of~\cite{FucGGKT-TACAS-09}.
}


\begin{abstract}
  Given a theory $T$ and two formulas $A$ and $B$ jointly unsatisfiable in $T$,
  a \emph{theory interpolant} of $A$ and $B$ is a formula $I$ such that (i) its
  non-theory symbols are shared by both $A$ and $B$, (ii) it is entailed by $A$ in
  $T$, and (iii) it is unsatisfiable with $B$ in $T$.  
  Theory interpolation has
  found several successful applications in model checking.
  We present a novel method for computing interpolants for
  ground formulas in the theory of equality. The method produces
  interpolants from colored congruence graphs representing derivations in
  that theory.
  These graphs can be produced by conventional congruence
  closure algorithms in a straightforward manner. By working with graphs, rather
  than at the level of individual proof steps, we are able to derive
  interpolants that are pleasingly simple (conjunctions of Horn clauses) and
  smaller than those generated by other tools.
  Our interpolation method can be seen as a theory-specific implementation of 
  a cooperative interpolation game between two provers. We present a generic 
  version of the interpolation game, parametrized by the theory $T$, and 
  define a general method to extract runs of the game from proofs in $T$ and 
  then generate interpolants from these runs.
  \end{abstract}

\maketitle



\section{Introduction}
\label{sec-introduction}

The \emph{Craig Interpolation Theorem}~\cite{Cra-JSL-57} asserts,
for every inconsistent pair of first-order formulas $A$, $B$,
the existence of a formula
$I$ that is implied by $A$, inconsistent with $B$, and written using only
logical symbols and symbols that occur in both $A$ and $B$.
Analogues of this result hold for a variety of logics and logic fragments.
Recently, they have found practical use in symbolic model checking.
Applications, starting with the work by McMillan~\cite{McC-CAV-03}, involve
computation of interpolants in propositional logic or in quantifier-free logics
with (combinations of) theories such as the theory of equality, linear rational
arithmetic, arrays, and finite
sets~\cite{McM-TCS-05,YorMus-CADE-05,KapMZ-SIGSOFT-06,CimGS-TACAS-08}.  
There are now several techniques that use interpolants to
obtain property-driven approximate reachability sets of transition relations,
or compute refinements for predicate abstraction~\cite{McM-TACAS-05,McM-CAV-06,JahMcM-CAV-05,JhaMcM-TACAS-06}. 

An important functionality in much of this work is the computation of ground
interpolants in the \emph{theory of equality}, also known as the theory of
\emph{uninterpreted functions} (\EUF).  
The ground interpolation algorithm for
this theory used in existing interpolation-based model checkers was developed by
McMillan \cite{McM-TCS-05}. It derives interpolants from proofs in a formal
system that contains rules for the basic properties of equality.

In this paper, 
which is a revised and expanded version of~\cite{FucGGKT-TACAS-09},
we present a novel method for ground \EUF interpolation. 
We compute interpolants from \emph{colored congruence graphs} that compactly
represent \EUF derivations from two sets of equalities, and can be produced in a
straightforward manner by conventional congruence closure algorithms,
as implemented in solvers for Satisfiability Modulo Theories 
(\eg, \cite{DefNS-JACM-05,NieOli-RTA-05}). 
Working
with graphs makes it possible to exploit the global structure of proofs 
to streamline interpolant generation. 
The generated interpolants are conjunctions of
Horn clauses, the simplest conceivable form for this theory. 
In most cases, they
are smaller and logically simpler than those produced by McMillan's method.

We restrict ourselves to input formulas $A$ and $B$ that 
are just conjunctions of \emph{literals}.
Such a restriction causes no loss of generality
because any interpolation procedure for conjunctions of literals
can be extended in a uniform way 
to arbitrary ground formulas--- and under the right conditions also combined 
with interpolation procedures for other theories~\cite{McM-TCS-05,CimGS-TACAS-08,GoeKT-CADE-09}.

Our interpolation method can be understood as the implementation of 
a cooperative interpolation game between two provers.
The game is not specific to the theory of equality and can be generalized
to other theories.
We present a general version of the interpolation game 
for a theory $\TT$ and 
define a generic method to extract runs of the game 
from \emph{local} refutations in $\TT$
and generate interpolants from these runs.

Our interpolation algorithm for \EUF is described and proved correct in \secref{sec-cc}.
In \secref{sec-examples}, we give a series of examples to highlight important
aspects of the algorithm.  A detailed comparison with McMillan's method is
given in \secref{sec-comparison}, together with experimental data on a set of
benchmarks derived from those in the SMT-LIB repository~\cite{BarST-SMTLIB}.
The general version of the interpolation game is described and proved correct
in \secref{sec-game}.

\subsection{Formal preliminaries}

We work in the context of first-order logic
with equality, and use standard notions of signature, 
term, literal, formula, clause, Horn clause, entailment, and so on.  
We use the symbol $=$ to denote the equality predicate in the logic as well as
equality at the meta-level, relying on context to disambiguate the
two.  For convenience, we treat all equations modulo symmetry, that
is, an equation of the form $s = t$ will stand indifferently for $s =
t$ or $t = s$.
For terms or formulas we will use ``ground'', \ie, variable-free, and 
``quantifier-free'' interchangeably since for our purposes
free variables can be always seen as free constants.

If $S, S_1, \ldots, S_n$ are sets of \defemph{sentences} 
(\ie, closed formulas) and $\varphi$ is a sentence, 
we write, as usual, $S_1, \ldots, S_n \models \varphi$ 
if $S_1 \cup \cdots \cup S_n$ logically entails $\varphi$; 
we write $S_1, \ldots, S_n \models S$ 
if $S_1, \ldots, S_n \models \psi$ for all $\psi \in S$.  
If $\TT$ is a theory, understood as a set of sentences,
we write $S \tentails \varphi$ as an abbreviation of 
$\TT \cup S \models \varphi$.  
We use the literals $\true$ and $\false$ as logical constants
denoting the universally true and the
universally false formula.  We say that a set of sentences $S$ is
\defemph{\TT-unsatisfiable} if $S \tentails \false$.

In FOL with equality, for any given signature $\Sigma$ the theory \EUF is
axiomatized by the empty set of sentences.  For convenience then, we
write $\models$ in place of $\models_\EUF$ and write ``unsatisfiable'' instead of
``\EUF-unsatisfiable'' when talking about that theory.  Also for
convenience, we do not distinguish a finite set of sentences from the
conjunction of its elements.


\section{Ground Theory Interpolation}

Interpolation is a property of \emph{logical fragments},
\ie, classes of formulas with an associated \emph{entailment} relation
over such formulas. 
To state it for a fragment \FF with entailment relation $\models_\FF$
we need know only a partition of the symbols used to build formulas in \FF 
into \emph{logical} and \emph{non-logical} symbols.

Let $\FF(X)$ be the set of all formulas in \FF whose non-logical symbols belong
to some set
$X$. By definition, $\FF$ has the \defemph{interpolation property} if for
every $A\in\FF(X)$ and $B\in\FF(Y)$ such that $A,B \models_\FF \false$, there exists
$I\in\FF(X\cap Y)$ such that $A\models_\FF I$ and $B,I\models_\FF \false$. The formula $I$ is an \defemph{(\FF-)interpolant} of $A$ and $B$. 
Note the asymmetry: $I$ is not an interpolant of $B$ and $A$; 
however, $\lnot I$ is---provided it belongs to $\FF$.

A classic theorem by William Craig~\cite{Cra-JSL-57} states that 
the fragment of all first-order logic formulas 
with the standard entailment relation has
the interpolation property. 
(The non-logical symbols are predicate and
function symbols, and \changed{free} variables.) 
The result also implies a \emph{modulo theory} generalization, 
where, for a given first-order theory \TT over a signature $\Sigma$,
the fragment $\FF$ is the set of all $\Sigma$-formulas
together with the entailment relation $\tentails$, 
and the symbols of $\Sigma$ are treated as logical. 
The case where \TT and $\Sigma$ are empty is Craig's original theorem.

Of particular interest is the interpolation property for quantifier-free
fragments of theories. The property may or may not hold, depending on the theory.  
Take, for example, 
the quantifier-free fragment of linear integer arithmetic, and 
let 
$A= \{x = 2y\}, B= \{x = 2z+1\}$.
The set $A \cup B$ is unsatisfiable in this theory, 
and the formula $\exists u.(x=2u)$ is an interpolant.
However, there is no quantifier-free interpolant for $A$ and $B$.

By definition, a theory has the \defemph{ground interpolation property} 
if its quantifier-free fragment has the interpolation property.  
Aside from \EUF,
several other theories of interest in model checking have this property,
including the theory of rational arithmetic among others~\cite{KapMZ-SIGSOFT-06,JaiCG-CAV-08}.

\begin{exa}
  The sets of inequalities $A=\{3x-z -2 \leq0,\, -2x+z-1\leq0\}$ and
  $B=\{3y-4z+12\leq0,\, -y+z -1\leq0\}$ are jointly unsatisfiable in the theory
  of rational arithmetic, as witnessed by
  the linear combination with positive coefficients 
\[
 2\cdot (3x-z -2 \leq 0) + 3\cdot
  (-2x+z-1 \leq0 ) + 1\cdot (3y-4z+12\leq 0) + 3\cdot (-y+z -1 \leq 0)
\]
which simplifies to
  $2\leq0$. The $A$-part of this linear combination $2\cdot (3x-z -2 \leq 0 ) + 3\cdot
  (-2x+z-1 \leq 0)$ gives us the interpolant $I = (z-7 \leq 0)$ for $A,B$.
  Generalizing what goes on in this example, one can obtain a ground
  interpolation procedure for the linear arithmetic with real coefficients. See,
  e.g., \cite{CimGS-TACAS-08}. \qed
\end{exa}

By the following lemma, if we want an algorithm for ground \TT-interpolation, 
it suffices to have one that works for inputs $A$ and $B$ 
that are \emph{sets of ground literals}. 

\begin{lem}
\label{lem-interpolant-reduction}
Let $\TT$ be a theory and suppose 
every pair of jointly $\TT$-unsatisfiable sets of literals
has a quantifier-free interpolant.
Then, $\TT$ has the ground interpolation property. 
\qed
\end{lem}

The reader is referred to \cite{McM-TCS-05,CimGS-TACAS-08,GoeKT-CADE-09} 
for effective proofs of the lemma---descriptions of 
a general mechanism to combine interpolation
procedures restricted to sets of literals with a method for computing
interpolants in propositional logic~\cite{Pud-JSL-97,McC-CAV-03}. 
With this justification, 
our interpolation method for \EUF focuses on sets of ground literals.


\section{Interpolation in \EUF}
\label{sec-examples}

\begin{figure}
\centering
\includegraphics[scale=0.6]{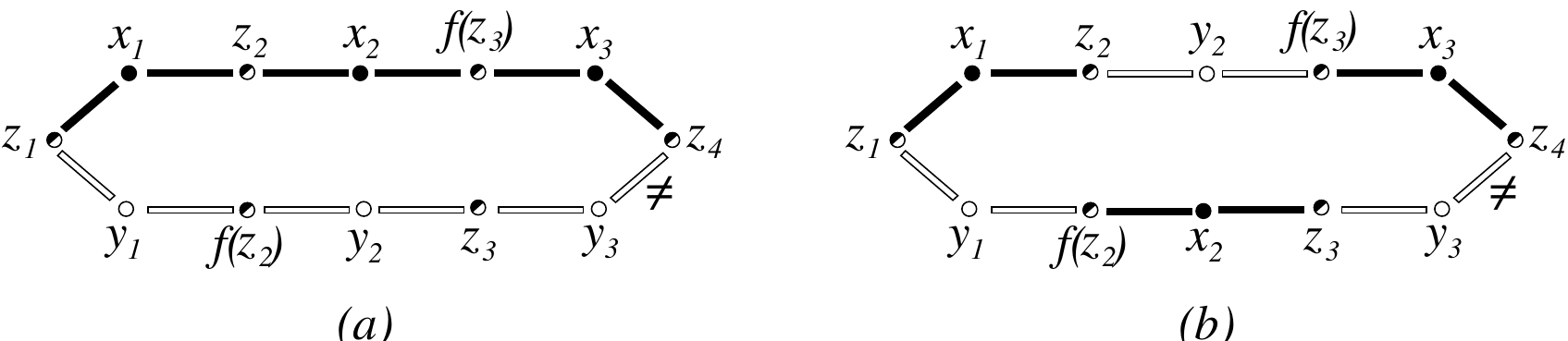}    
\caption{
Solid (hollow) edges represent literals from the set $A$ (set $B$) in  
Example~\ref{ex-transitivity}.
For the vertex coloring convention, see \exref{ex-coloring}.
}
\label{fig-transitivity}
\end{figure}

It is instructive to look first at some examples of interpolants
for pairs of literal sets
$A$ and $B$ jointly unsatisfiable in \EUF.

\begin{exa}
\label{ex-transitivity}
The picture in \figref{fig-transitivity}(a) demonstrates the joint unsatisfiability of
\[
\begin{array}{lll}
A & = & 
\{z_1 = x_1, \, 
  x_1 = z_2, \, 
  z_2 = x_2, \,
  x_2 = f(z_3), \, 
  f(z_3) = x_3, \,
  x_3 = z_4,\,
  f(z_2) = x_2,\,
  x_2 = z_3
\}, \\
B & = & 
\{z_1 = y_1, \,
  y_1 = f(z_2), \,
  f(z_2) = y_2, \, 
  y_2 = z_3, \, 
  z_3 = y_3, \, 
  z_2 = y_2,\,
  y_2 = f(z_3),\,
  y_3 \neq z_4 
\}
\end{array}
\]
which follows by the transitivity of equality.  
An interpolant is the equality $z_1=z_4$ that summarizes the transitivity 
$A$-chain in the figure. 
For the variation in \figref{fig-transitivity}(b), 
which provides an alternative demonstration of the joint unsatisfiability
of $A$ and $B$,
an interpolant is the conjunction
$z_1=z_2 \land f(z_3)=z_4 \land f(z_2)=z_3$ of summaries of $A$-chains. 

For yet another variation,
this time with slightly different sets $A$ and $B$,
modify \figref{fig-transitivity}(b) by moving the disequality
sign to the edge $\langle x_3,z_4\rangle$.  
There, an interpolant is $z_1=z_2 \land f(z_3)\neq z_4 \land f(z_2)=z_3$. 
\qed
\end{exa}

\begin{exa}
\label{ex-first-horn}
When the unsatisfiability of $A\cup B$ involves the congruence property of $=$, an interpolant
in the form of a conjunction of equalities need not exist.  Let 
\[
\begin{array}{lll}
A = \{u_1=x\cdot u_0,\, v_1=x\cdot v_0\}
& \text{ and } &
B = \{u_0 = v_0,\, u_1\neq v_1\}
\end{array}
\] 
where the dot is an infix binary function symbol.
There are no equalities entailed by $A$ that do not contain
$x$. The transitivity chain $u_1 = x \cdot u_0 = x\cdot v_0 = v_1$ contradicts
$u_1\neq v_1\in B$, but its middle equality is not entailed by $A$. 
 However, $A$ does entail it under the condition $u_0=v_0$ that $B$ provides. That gives us the interpolant $u_0=v_0\limplies u_1=v_1$. 
\end{exa}

\begin{figure}[htbp]
\centering
\includegraphics[scale=0.6]{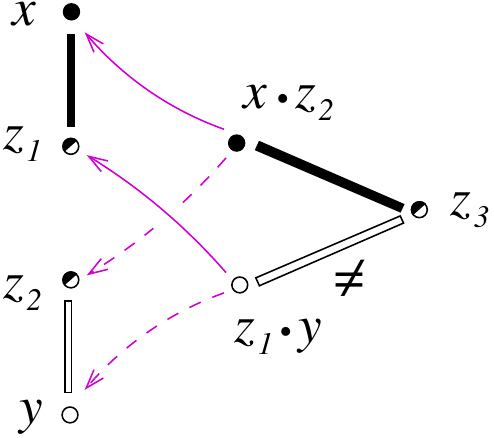}
\caption{
The solid and the dashed arrows point to the two equalities
of $A \cup B$ that 
entail the equality $x\cdot z_2=z_1\cdot y$. (\exref{ex-uncolorable})}
\label{fig-uncolorable}
\end{figure}

\begin{exa}\label{ex-uncolorable}
With
\[
\begin{array}{lllllll}
A & = & \{x=z_1,\, x\cdot z_2=z_3\} 
& \text{ and } &
B & = & \{y=z_2,\, z_1\cdot y\neq z_3\}
\end{array}
\]
pictured in \figref{fig-uncolorable}, 
we can derive $\false$ from the chain 
$z_3 = x \cdot z_2 = z_1\cdot y \neq z_3$, 
where the congruence reasoning that produces the
middle equality $x \cdot z_2 = z_1\cdot y$ uses an equality 
from $A$ ($x=z_1$) and an equality from $B$ ($z_2=y$), 
and cannot be derived from either $A$ or $B$ alone. 
A simple split of the problematic equality into two produces 
a chain in which every literal follows from either $A$ or $B$: 
$z_3 = x \cdot z_2 = z_1\cdot z_2 = z_1\cdot y \neq z_3$.
The summary $z_3 = z_1\cdot z_2$ of the $A$-chain is 
and interpolant of $A$ and $B$.
The upshot here is that creating an interpolant may require terms 
(in this case, $z_1\cdot z_2$)
that do not occur in either $A$ or $B$. 
See \lemref{lem-colorable-graph} below.
\qed
\end{exa}

\section{Interpolants From Congruence Closure}
\label{sec-cc}

Efficient decision procedures 
for the satisfiability of sets of literals in EUF 
are typically based on \emph{congruence closure}~\cite{NelOpp-JACM-80,DefNS-JACM-05,NieOli-RTA-05}.
In this section,
we show that one can minimally modify such procedures
to produce interpolants as well.

\subsection{Congruence Closure}
\label{sec-cca}

The \defemph{congruence closure algorithm} takes as inputs 
\begin{iteMize}{$\bullet$}
\item
a finite set $E$ of ground equalities and
\item
a finite subterm-closed set $T$ of ground terms. 
\end{iteMize}
Its state is an \emph{undirected} graph $\G$, 
initialized so that its
vertex set is $T$ and its edge set is empty.  
We write $u\sim v$ to mean that $u$ and $v$ are connected by a path in $\G$.
The algorithm proceeds as follows.

\ 

\begin{iteMize}{(\textbf{cc0})}
\item [\cczero]
Let $\G = (T,\emptyset)$
\item [\ccone] 
Choose distinct $s,t\in T$ such that $s\not \sim t$ and either 
\begin{iteMize}{(\bf a)}
  \item  $(s=t)\in E$; or
  \item  $s$ is $f(s_1,\ldots,s_k)$, $t$ is $f(t_1,\ldots,t_k)$, and $s_1
    \sim t_1$, \ldots, $s_k\sim t_k$.
  \end{iteMize}
Then  add the edge $\langle s,t\rangle$ to $\G$
\item [\cctwo] 
Repeat {\ccone} for as long as possible.
 \end{iteMize}

\begin{thm}
\label{thm-cc}
\cite{NelOpp-JACM-80,NieOli-RTA-05} 
Let $\sim$ be the equivalence relation obtained by running 
the congruence closure algorithm above. 
For every $s,t\in T$, one has $E\models s=t$ if and only if $s\sim t$.
Moreover, the set $E\cup\{s\neq t\;|\; s\not \sim t\}$ is satisfiable. \qed
\end{thm}

If $L$ is an arbitrary set of ground \EUF literals,
let $L = L_= \cup L_{\neq}$,
where $L_=$ and $L_{\neq}$ consists respectively of 
the equalities and disequalities of $L$.
To check whether $L$ is satisfiable, it suffices to run the
congruence closure algorithm with $E=L_=$ and $T$ consisting of all the terms
(and subterms) occurring in $L$. By \thmref{thm-cc}, $L$ is satisfiable if and only if
$s\not\sim t$ holds for every disequality $s\neq t$ in $L_{\neq}$. 
Conversely,
$L$ is unsatisfiable if and only if $L_=\cup\{\delta\}$ is unsatisfiable 
for some $\delta\in L_{\neq}$.

\subsection{Congruence Graphs }
\label{sec-cg}

For any finite set $E$ of ground equalities and
a finite subterm-closed set $T$ of ground terms,
a \defemph{congruence graph} over $E$ and $T$ is any intermediate graph $\G$
obtainable by the congruence closure algorithm above. 
We will not mention the term set $T$ when it is understood or unimportant.

The assumption $s\not\sim t$ in Step~\ccone\ ensures that 
every congruence graph is acyclic.  
Thus, if $u\sim v$ in a congruence graph $\G$, 
there is a unique \defemph{path} connecting them.
We denote this path by $\path u v$. 
\defemph{Empty paths} are those of the form $\path u u$.

We call an edge of a congruence graph $\G$
\defemph{basic} or \defemph{derived} depending 
on whether it has been introduced in $\G$ respectively
because of Condition (\textbf{a}) or Condition (\textbf{b}) of Step~\ccone.
A derived edge $\langle f(u_1,\ldots,u_k),f(v_1,\ldots,v_k)\rangle$ has $k$
\defemph{parent paths} $\path{u_1}{v_1}$,\ldots,$\path{u_k}{v_k}$, 
some (but not all) of which may be empty.

\begin{exa}
Each of the graphs in Figures \ref{fig-transitivity} and \ref{fig-uncolorable},
\emph{when we delete from it the edge marked with the $\neq$ symbol}, 
is a congruence graph over the corresponding set of equalities $(A\cup B)_=$. 
All edges in these graphs are basic; 
in Figure~\ref{fig-uncolorable}, 
a derived edge between the nodes $x \cdot z_2$ and $z_1 \cdot y$
could be added as a consequence of the basic edges pointed to by the arrows.
\qed
\end{exa}

\begin{exa}
\label{ex-mainex}
Let $E=(A\cup B)_=$ where
\[
\begin{array}{lll}
 A & = & \{x_1=z_1,\, z_2=x_2,\, z_3=f\,x_1,\, f\,x_2=z_4,\, x_3=z_5,\,
           z_6=x_4,\, z_7=f\,x_3,\, f\,x_4=z_8\},
 \\
 B & = & \{z_1=z_2,\, z_5=f\,z_3,\, f\,z_4=z_6,\, y_1=z_7,\, z_8=y_2,\, y_1\neq y_2\} \ .  
\end{array}
\]
\figref{fig-mainex}(b) depicts a congruence graph over $E$. 
The basic edges are shown in \figref{fig-mainex}(a); 
each corresponds to an equality in $E$.
Since $f$ is unary, each of the three derived edges has one parent path.
\qed
\end{exa}

\subsection{Colorable Congruence Graphs}
\label{sec-ccg}

Let $A$ and $B$ be sets of ground literals and 
let $\Sigma_A$ and $\Sigma_B$ be the sets of non-logical 
symbols that occur in $A$ and $B$, respectively. 
Terms, literals, and formulas over $\Sigma_A$ will be called \defemph{$A$-colorable},
those over $\Sigma_B$ will be called \defemph{$B$-colorable}.
Such expressions will be called \defemph{colorable} 
if they are either $A$-colorable or $B$-colorable,
and \defemph{$AB$-colorable} if they are both.

\begin{exa}
In \exref{ex-uncolorable}, 
$\Sigma_A=\{x,z_1,z_2,z_3,\cdot\}$ and $\Sigma_B=\{y,z_1,z_2,z_3,\cdot\}$. 
Terms and equalities without occurrences of either $x$ or $y$ are 
$AB$-colorable. 
The term $x\cdot y$ and the equality $x\cdot z_2 = z_1 \cdot y$ are 
not colorable.
\qed
\end{exa}

We extend the above definitions to edges of congruence graphs over $A\cup B$ 
so that
an edge $\langle s,t\rangle$ has the same colorability attributes as the equality $s=t$.
Note that basic edges are always colorable. 
Finally, 
we define a path in a congruence graph (resp., a congruence graph) 
to be \defemph{colorable} if all edges in the path (resp., graph) are colorable.

\begin{exa}
\label{ex-coloring}
The congruence graphs derived from graphs in Figures
\ref{fig-transitivity} and \ref{fig-uncolorable} by removing 
their disequality edges are all colorable.  
Among the vertices (which are terms),
the half-filled ones are $AB$-colorable,
the dark ones are $A$-colorable but not $B$-colorable, 
and the light ones are $B$- but not $A$-colorable;
however, if we add the derived edge $\langle x\cdot z_2,z_1 \cdot y\rangle$ 
to the graph in \figref{fig-uncolorable}, 
it will not be colorable.
\qed
\end{exa}

For our purposes, 
the uncolorability of some congruence graphs is not a problem 
thanks to the following result. 

\begin{lem}
 \label{lem-colorable-graph} 
If $s$ and $t$ are colorable terms and if $A,B \models s=t$, 
then there exist a term set $T$ and
a colorable congruence graph over $(A\cup B)_=$ and $T$ in which $s\sim t$.
\end{lem}
\proof
 This is essentially Lemma 2 of \cite{YorMus-CADE-05}, and the
  proof is constructive. Start with any congruence graph $\G$ with colorable
  vertices in which $s\sim t$ holds. If there are uncolorable edges, let $e=\langle
  f(u_1,\ldots,u_k),f(v_1,\ldots,v_k)\rangle$ be a minimal such edge in the derivation order.
  Thus, the parent paths $\path{u_i}{v_i}$ are all colorable, and each of them
  connects an $A$-colorable vertex with a $B$-colorable one. It follows that
  there exists an $AB$-colorable vertex $w_i$ on each path
  $\path{u_i}{v_i}$ (which may be one of its endpoints).  The term
  $f(w_1,\ldots,w_k)$ is $AB$-colorable, so  
  add it to the vertex set of $\G$ and
  replace $e$ in $\G$ with the two
  edges $\langle f(u_1,\ldots,u_k),f(w_1,\ldots,w_k)\rangle$ and $\langle
  f(w_1,\ldots,w_k),f(v_1,\ldots,v_k)\rangle$, both of which are colorable. Now repeat the
  process until all uncolorable edges of $\G$ are eliminated. 
  The set $T$ is the final set of vertices of $\G$.
  \qed
\bigskip

Note that the proof of Lemma~\ref{lem-colorable-graph} 
provides an effective procedure
for turning any uncolorable graph into a colorable one.
Using a data structure for the congruence graph that also
maintains for each derived edge a pointer to its parent paths
allows a linear-time bottom-up implementation of the procedure.

\begin{exa}
Consider again the uncolorable congruence graph
obtained by adding the derived edge $\langle x\cdot z_2,z_1 \cdot y\rangle$ 
to the graph in \figref{fig-uncolorable}.
Using the procedure in the proof of Lemma~\ref{lem-colorable-graph}
we can turn it into a colorable congruence graph by replacing the edge
$\langle x\cdot z_2,z_1 \cdot y\rangle$
with the edges
$\langle x\cdot z_2, z_1 \cdot z_2\rangle$
and
$\langle z_1 \cdot z_2, z_1 \cdot y\rangle$.
\qed
\end{exa}

\subsection{Colored Congruence Graphs}
\label{sec-coloredcg}

\begin{figure}
  \centering
\includegraphics[scale=0.57]{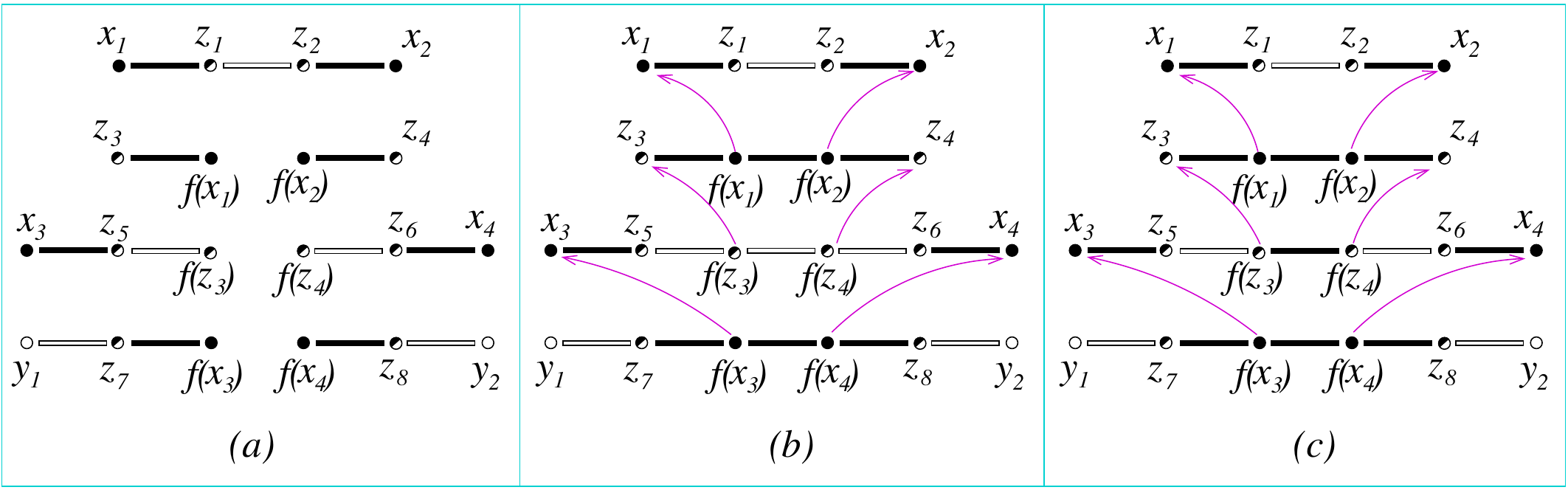}  
  \caption{Congruence graphs over $(A\cup B)_=$, with  $A$ and $B$ from
    \exref{ex-mainex}. The connection between a derived edge and its parent is
    indicated by a pair of arrows.}
\label{fig-mainex}  
\end{figure}

Assume (without loss of generality) that the literal sets $A,B$ are disjoint. A
\defemph{coloring} of a colorable congruence graph over $(A\cup B)_=$ is an
assignment of a unique color $A$ or $B$ to each edge of the graph, such that
\begin{iteMize}{$\bullet$}
\item basic edges are assigned the color of the set they belong to,
\item every edge colored $X$ has both endpoints $X$-colorable ($X\in\{A,B\}$).
\end{iteMize}
Thus, to color a colorable congruence graph, the only choice we have is with
$AB$-colorable derived edges, and each of them can be colored arbitrarily.  
In
the terminology of the interpolation game described later in \secref{sec-game}, this means
choosing which prover derives an $AB$-equality in a situation when either of
them could do it. In \figref{fig-mainex}(b,c) we have two colored congruence
graphs. They differ only in the coloring of $\langle f(z_3),f(z_4)\rangle$---the only
derived edge with $AB$-colorable endpoints.

In a colored graph, we can speak of \defemph{$A$-paths} (whose edges are all
colored $A$), and \defemph{$B$-paths}. There is also a color-induced
factorization of arbitrary paths, where a \defemph{factor} of a path $\pi$ is a
maximal subpath of $\pi$ consisting of equally colored edges.
Clearly, every path can be uniquely represented as a concatenation of its
factors, the consecutive factors having distinct colors.

\subsection{The Interpolation Algorithm}
\label{sec-interpolation}

Our goal is to construct an interpolant for the pair of sets $A$ and $B$ 
of ground literals that are jointly inconsistent in \EUF.  
The algorithm presented below relies on the results 
in the previous subsection which guaranteed the existence
(and computability)
of a disequality $s\neq t$ in $A\cup B$ and 
a colored congruence graph $G$ over $(A\cup B)_=$ 
such that $s$ and $t$ are connected in $G$.

A path $\path u v$ in a congruence graph represents 
the equality $u=v$ between its endpoints,
summarizing the reflexivity, symmetry and transitivity inferences
encoded by the path.
The algorithm presented below builds an interpolant 
as a conjunction of Horn clauses 
whose atoms are $AB$-colorable equalities, 
each summarizing an A-path or a B-path of the graph $G$.
The algorithm minimizes the number of such equalities
by breaking paths only along their color-induced factorization
(as opposed to other, finer partitions).

We will write $\eq{\pi}$ to denote the equality represented by the path $\pi$. 
More generally, if $P$ is a set of paths, $\eqs P$ is the
corresponding set of equalities.
For convenience, 
we will take $\eq{\path u u}$ to be $\true$, instead of $u = u$, for each empty path $\path u u$.
(Similarly for $\eqs P$, when $P = \emptyset$.)

For every path $\pi$ in a colored congruence graph $G$, 
we define below the associated \defemph{$B$-premise set} $\BB(\pi)$, 
the \defemph{$A$-justification} $J(\pi)$,
and the \defemph{path interpolant} $I(\pi)$.
Intuitively,
for an $A$-path $\pi$,
the $B$-premise set collects all the maximal $B$-paths in $G$ 
that allow the construction of $\pi$ 
(by connecting ancestors of edges in $\pi$);
the $A$-justification is an implication from
all the equalities represented by $\pi$'s $B$-premises to $\eq \pi$,
capturing the fact that $\eq \pi$ is a consequence of $A$ 
and all those equalities;
the path interpolant is 
the conjunction of $\pi$'s $A$-justification
together with the path interpolants for each of its $B$-premises.
For a $B$-path $\pi$,
the $B$-premise set is simply $\{\pi\}$;
the $A$-justification is, trivially, $\eq{\pi} \limplies \eq{\pi}$ 
(and actually never used);
the path interpolant is the conjunction of all the path interpolants
of $\pi$'s parent paths.

For instance, for the congruence graph in Figure~\ref{fig-mainex}(b),
$\BB(\path{z_3}{z_4}) = \{\path{z_1}{z_2}\}$, 
$J(\path{z_3}{z_4}) = (z_1 = z_2 \limplies z_3 = z_4)$, 
$I(\path{z_5}{z_6}) = I(\path{z_3}{z_4}) = (z_1 = z_2 \limplies z_3 = z_4)$,
$\BB(\path{z_7}{z_8}) = \{\path{z_5}{z_6}\}$, 
$J(\path{z_7}{z_8}) = (z_5 = z_6 \limplies z_7 = z_8)$, 
and
$I(\path{z_7}{z_8}) = (z_5 = z_6 \limplies z_7 = z_8) \land
                      (z_1 = z_2 \limplies z_3 = z_4)$.

\begin{align}
\label{eq-Bdef}
\BB(\pi) & = 
\begin{cases}
 \bigcup\{\BB(\sigma) \mid \text{$\sigma$ is a factor of $\pi$}\}
 & \text{if $\pi$ has $\geq2$ factors}
 \\
 \{\pi\} & \text{if $\pi$ is a $B$-path}
 \\
 \bigcup\{\BB(\sigma) \mid \text{$\sigma$ is a parent of an edge of $\pi$}\}
 & \mbox{if $\pi$ is an $A$-path}
\end{cases}
\\[2ex]
J(\pi) & = (\bigwedge\eq{\BB(\pi)}) \limplies \eq{\pi}
\\[2ex]
I(\pi) & = 
\begin{cases}
 \bigwedge\{I(\sigma) \mid \text{$\sigma$ is a factor of $\pi$}\}
 & \text{if $\pi$ has $\geq2$ factors}
 \\
 \bigwedge\{I(\sigma) \mid \text{$\sigma$ is a parent of an edge of $\pi$}\}
  & \mbox{if $\pi$ is a $B$-path}
 \\
 J(\pi) \;\land \; \bigwedge\{I(\sigma) \mid \sigma\in\BB(\pi)\}
 & \text{if $\pi$ is an $A$-path}
\end{cases}
\end{align}
\medskip

Empty parent paths $\sigma$ in the definitions of $\BB(\pi)$ and $I(\pi)$ can be
ignored because $\eq{\sigma} = J(\sigma) = I(\sigma) = \true$ when $\sigma$ is empty.

We also need a modified interpolant function $I'$, expressed in terms of $I$ as
follows.  The argument path $\pi$ is first decomposed as $\pi = \pi_1\theta\pi_2$, where
$\theta$ is the largest subpath with $B$-colorable endpoints, or an empty path if
there are no $B$-colorable vertices on $\pi$; then 
\begin{align}
I'(\pi) = \textstyle{I(\theta) \ \land \ \bigwedge\{I(\tau)\,|\, \tau\in\BB(\pi_1)\cup\BB(\pi_2)\}}
\textstyle{\ \land \ \big( \bigwedge\eq{\BB(\pi_1) \cup \BB(\pi_2)} \ \limplies\ \lnot\eq{\theta}\big)}
\label{eq-Iprimedef}
\end{align}

It is not difficult to see that 
$\BB, J, I$ and $I'$ are all well defined and computable.
In particular,
$I'$ is well defined because $\pi_1,\theta,\pi_2$ are uniquely determined by $\pi$ if
$\theta$ is not empty, and if $\theta$ is empty, the way we write $\pi$ as $\pi_1\pi_2$ is
irrelevant. Note that when $\pi=\theta$, we have $I'(\pi) = I(\pi) \land \lnot\eq{\pi}$. 
\medskip

The \EUF \defemph{ground interpolation algorithm}, given as input two jointly
inconsistent (disjoint) sets $A,B$ of literals, proceeds as follows.
\medskip

\begin{iteMize}{(\textbf{i1})}
\item Run the congruence closure algorithm to find a congruence graph
  $\G$ over $(A\cup B)_=$ and a disequality $(s\neq t)\in A\cup B$ such that $s\sim t$ in $\G$
  [\secref{sec-cca},\secref{sec-cg}]. 
\item  Modify $\G$ as necessary to make it colorable
  [\secref{sec-ccg}], then color it [\secref{sec-coloredcg}].
\item If $(s\neq t)\in B$, return  $I(\path s t)$; if $(s\neq t)\in A$,
  return $I'(\path st)$.
\end{iteMize}
\medskip

\def\bpath#1#2#3#4{\BB(\path {{#1}_{#2}}{{#3}_{#4}})}
\def\ipath#1#2#3#4{I(\path {{#1}_{#2}}{{#3}_{#4}})}
\def\ppath#1#2#3#4{\path {{#1}_{#2}}{{#3}_{#4}}}

\begin{exa}
\label{ex-run}
Let us run the algorithm for $A,B$ in \exref{ex-mainex}, using the colored
congruence graph in \figref{fig-mainex}(b). Since $y_1\neq y_2\in B$, the
interpolant is computed by applying $I$ to $\path{y_1}{y_2}$:
\begin{align*}
\ipath y1y2  
 & = \ipath y1z7 \land \ipath z7z8 \land \ipath z8y2
   =  \true \land \ipath z7z8 \land \true \\
 & = \ipath z7z8
   = J(\ppath z7z8) \land \bigwedge\{I(\sigma) \mid \sigma\in\bpath z7z8\}
\end{align*}

\noindent In turn,
\(
\bpath z7z8  
   = \bpath x3x4
   = \bpath x3z5 \cup\bpath z5z6 \cup \bpath z6x4
   = \emptyset \cup\{\ppath z5z6\} \cup\emptyset
   = \{\ppath z5z6\}.
\)
Thus, $J(\ppath z7z8) = (z_5=z_6\limplies z_7=z_8)$. 
Continuing the main computation: 
\begin{align*}
\ipath y1y2  
 & = J(\ppath z7z8) \land \ipath z5z6 \\
 & = J(\ppath z7z8) \land 
     J(\ppath z3z4) \land \bigwedge\{I(\sigma) \mid \sigma\in\bpath z3z4\}
\end{align*}

\noindent Now, 
\(
\bpath z3z4
 = \BB(\path{x_1}{x_2})
 = \BB(\path{x_1}{z_1}) \cup\BB(\path {z_1}{z_2}) \cup \BB(\path{z_2}{x_2})
 = \emptyset \cup\{\path{z_1}{z_2}\}\cup\emptyset =\{\ppath z1z2\}.
\)
Thus, $J(\ppath z3z4) = (z_1=z_2\limplies z_3=z_4)$. 
Back to the main computation again, 
\begin{align*}
\ipath y1y2 
 & = J(\ppath z7z8) \land J(\ppath z3z4) \land \ipath z1z2
   = J(\ppath z7z8) \land J(\ppath z3z4)\land\true \\
 & = (z_5=z_6\limplies z_7=z_8) \land (z_1=z_2\limplies z_3=z_4)
\end{align*}

The reader can verify that using the graph in
  \figref{fig-mainex}(c) results in a different interpolant: 
\begin{align*}
I(\path{y_1}{y_2})
 & = (z_5=f(z_3) \land z_6=f(z_4)\land z_1=z_2) \limplies z_7=z_8 .
\end{align*}
\end{exa}

\subsection{Correctness }
\label{sec-proof}

Our main correctness results can be expressed as follows.

\begin{thm}
\label{thm-correctness}
With any jointly inconsistent sets $A$,$B$ of \EUF literals as inputs, the \EUF
ground interpolation algorithm (\secref{sec-interpolation}) terminates and
returns an interpolant for $A$,$B$ that is a conjunction of Horn clauses.
\qed
\end{thm}

To prove the theorem we need to introduce some additional notions and notation.
For the rest of the section let $G$ be a colored congruence graph.

The termination of our recursive definitions and other inductive arguments 
are proved using a well-founded relation $\prec$ over paths of $G$.  
Define $\sigma\prec_1\pi$ to hold whenever:
\begin{iteMize}{$\bullet$}
\item  $\pi$ has more than one factor and $\sigma$ is one of them, or
\item  $\sigma$ is a parent path of an edge of $\pi$.
\end{iteMize}
Then, define $\prec$ as the transitive closure of $\prec_1$.
It is not difficult to see that the relation $\prec$ is well-founded.  
Note that minimal elements under $\prec$ are
the paths all of whose edges are basic and of the same color.

The following equations redefine the set $\BB(\pi)$ of $B$-premises and introduce
the analogous set $\AA(\pi)$ of \defemph{$A$-premises}.
\begin{gather} \label{eq-Adef}
\AA(\pi) \; =\;  \textstyle{\{\mbox{$A$-factors of $\pi$}\} \  \cup
\  \AA(\{\mbox{parent paths of $B$-edges of $\pi$}\})}
\\ 
\label{eq-Bdeff}
\BB(\pi)  \; =\;  \textstyle{\{\mbox{$B$-factors of $\pi$}\} \  \cup
\ \BB(\{\mbox{parent paths of $A$-edges of $\pi$}\})}   
\end{gather}

Here and in the sequel, we use the convention $f(P)=\bigcup\{f(\sigma)\,|\,\sigma\in P\}$ for
extending a set-valued function $f$ defined on paths to a function defined on
sets of paths.  Observe that \eqref{eq-Bdeff} is just a restatement of
\eqref{eq-Bdef}.  Also, the arguments in the recursive calls are smaller than
$\pi$ under the relation $\prec$, so termination is guaranteed. 

The basic properties of $\AA$ are collected in the following lemma. 
The analogous properties of $\BB$ follow by symmetry.

\begin{lem}
  \label{lem-monotonicity}
  Let $\pi$ be an arbitrary non-empty path in $\G$.
   \begin{enumerate}[\em(1)]
  \item If $\pi$ is an $A$-path, then $\AA(\pi) = \{\pi\}$; otherwise, $\sigma\prec\pi$
    for every $\sigma\in\AA(\pi)$. 
  \item  If $\sigma\in\AA(\pi)$, then $\AA(\sigma) \subseteq\AA(\pi)$.
  \item  If the endpoints of $\pi$ are $B$-colorable, then the endpoints
    of all paths in $\AA(\pi)$ are $AB$-colorable.
  \end{enumerate}
\end{lem}

\proof
  All three parts are proved by well-founded induction.  
  
  \emph{(1)} If $\pi$ is an $A$-colored path, then $\pi$ is the only element of
  $\AA(\pi)$ (by definition).  If $\pi$ is not an $A$-colored path and $\tau$ is an
  element of $\AA(\pi)$, then $\tau$ is either an $A$-factor of $\pi$ and so $\tau\prec\pi$
  holds, or $\tau\in\AA(\sigma)$ for some parent $\sigma$ of a $B$-edge of $\pi$. In the
  latter case, $\tau\prec\pi$ holds because of $\sigma\prec\pi$ and the consequence $\tau\preceq\sigma$
  of the induction hypothesis.
  
  \emph{(2)} If $\sigma$ is an $A$-factor of $\pi$, then $\AA(\sigma)=\{\sigma\}\subseteq\AA(\pi)$. If
  $\sigma\in\AA(\tau)$ where $\tau$ is a parent path of a $B$-edge of $\pi$, then
  $\AA(\sigma)\subseteq\AA(\tau)\subseteq\AA(\pi)$, the first inclusion by induction hypothesis, the second
  from the definition of $\AA$.

  \emph{(3)} Since parent paths of any $B$-edge must have $B$-colorable
  endpoints, for the inductive argument we only need to check that every
  $A$-factor of a path $\pi$ with $B$-colorable endpoints has $AB$-colorable
  endpoints.  Indeed, $A$-colorability of endpoints of $A$-factors is obvious.
  For $B$-colorability, observe that an endpoint of an $A$-factor of $\pi$ is
  either also an endpoint of a $B$-factor of $\pi$, or an endpoint of $\pi$
  itself.  \qed
\bigskip

The following lemma justifies the names \emph{$A$-premises} and
\emph{$B$-premises}. 
Intuitively,
$B$-premises are the $B$-paths whose summaries, together with $A$, 
entail $\eq{\pi}$.
Dually,
$A$-premises are the $A$-paths whose summaries, together with $B$, 
entail $\eq{\pi}$.

\begin{lem}
\label{lem-premises}
\(
 A, \eqs{\BB(\pi)}\models \eqs \pi
 \text{ and }
 B, \eqs{\AA(\pi)} \models \eqs \pi
\)
for every path $\pi$ in $\G$.
\end{lem}

\proof
  We prove the first claim only, by well-founded induction based on $\prec$.
  Viewing $\pi$ as the concatenation of its $B$-factors and $A$-edges, we have by
  transitivity
  $$\eqs{\mbox{$B$-factors of $\pi$}}, \; \eqs{\mbox{$A$-edges of $\pi$}} \models \eqs
  \pi$$
  and then, since $A\models\eq{ e}$ for every basic $A$-edge $e$ (by definition of
  edge coloring),
  $$A,\, \eqs{\mbox{$B$-factors of $\pi$}}, \; \eqs{\mbox{derived
      $A$-edges of $\pi$}} \models\eqs \pi.$$
  For every derived edge
  $e$ we have $\eqs{\mbox{parents of $e$}} \models \eqs e$. Thus,
  $$A,\, \eqs{\mbox{$B$-factors of $\pi$}}, \; \eqs{\mbox{parents of $A$-edges of
      $\pi$}} \models \eqs \pi,$$
so it suffices to prove 
$A, \ \eqs{\BB(\pi)} \models \eqs \sigma$ for every $\sigma$ that is either a $B$-factor of
$\pi$ or a parent of an $A$-edge of $\pi$. 

In the first case, the claim holds since $\sigma\in\BB(\pi)$. 
In the second case, we have $\sigma\prec\pi$, so the
induction hypothesis gives us $A, \, \eqs{\BB(\sigma)} \models \eqs \sigma$. 
To finish the proof, just use the fact $\BB(\sigma)\subseteq\BB(\pi)$,
by \lemref{lem-monotonicity}(ii).
\qed
\bigskip

Define the \defemph{cumulative set of premises}  (cf.~\secref{sec-game})
of a path $\pi$ as
\begin{equation}
  \label{eq-cumulative}
  \prem \pi \; =\; \{\pi\} \, \cup\, \prem{\BB(\AA(\pi))}\ .
\end{equation}
The termination of this
recursive definition follows from \lemref{lem-monotonicity}(i).

\begin{lem}
  \label{lem-Iequivalence}
For every path $\pi$ in $\G$, \ 
$I(\pi) \:=\: \bigwedge\,\{J(\sigma) \,|\, \sigma\in\AA(\prem \pi)\}$.
\end{lem}

\def\pprem#1{\PP'(#1)}

\proof
Let $\pprem \pi = \AA(\PP(\pi))$. 
From \eqref{eq-cumulative}, we have
\begin{equation}
\label{eq-pp}
\pprem \pi \; =\; \AA(\pi) \, \cup\, \pprem{\BB(\AA(\pi))}   
\end{equation}
It suffices to check that 
\begin{eqnarray*}
\label{eq-pprem}
\pprem \pi & = & 
\begin{cases}
 \bigcup\{\pprem \sigma \mid \mbox{$\sigma$ is a factor of $\pi$}\}
  & \text{if $\pi$ has $\geq2$ factors} 
 \\
 \bigcup\{\pprem \sigma \mid \mbox{$\sigma$ is a parent of an edge of $\pi$}\}
  & \text{if $\pi$ is a $B$-path}
 \\
 \{\pi\} \;\cup\; \bigcup\{\pprem \sigma \mid \sigma\in\BB(\pi)\}
  & \text{if $\pi$ is an $A$-path}
\end{cases}
\end{eqnarray*}
  
For the first case, 
suppose $\pi=\pi_1\cdots\pi_k$ is the factorization of $\pi$. 
By definition of $\AA$, 
we have $\AA(\pi) = \AA(\pi_1) \cup\cdots\cup\AA(\pi_k)$. 
The desired equality 
$\pprem\pi = \pprem{\pi_1}\cup\cdots\cup \pprem{\pi_k}$ 
then follows from \eqref{eq-pp}. 
  
Assume now that $\pi=e_1\cdots e_k$ is a $B$-path. 
By definition of $\AA$,
we have $\AA(\pi) = \AA(E_1) \cup \cdots \cup \AA(E_k)$, 
where $E_i$ is the set of parent paths of the edge $e_i$ 
for $i=1,\ldots,k$. 
Again, the desired equality 
$\pprem \pi = \pprem{E_1} \cup \cdots \cup \pprem{E_k}$ 
follows from \eqref{eq-pp}. 
  
Finally, assume that $\pi$ is an $A$-path.
Now $\AA(\pi) = \{\pi\}$ and so $\pprem \pi = \{\pi\} \cup \pprem {\BB(\pi)}$,
again by \eqref{eq-pp}.
\qed
\bigskip

\begin{lem}
\label{lem-main}
$B, I(\pi) \models \eqs \pi$
for every path $\pi$ in $\G$ with $B$-colorable endpoints.
\end{lem}
\proof
  We argue by induction along $\prec$. Let $\sigma$ be an arbitrary $A$-premise of $\pi$
  and $\tau$ an arbitrary $B$-premise of $\sigma$.  The endpoints of $\tau$ are
  $B$-colorable, because in general, every $B$-premise of any path is a
  $B$-factor of some path, and every $B$-factor of any path has $B$-colorable
  endpoints. Thus, the induction hypothesis applies to $\tau$ and we have
  $B,I(\tau)\models \eq{\tau}$. From equation \eqref{eq-pp} we have $\pprem \tau \subseteq\pprem
  \pi$, so we can derive $I(\pi) \models I(\tau)$ using \lemref{lem-Iequivalence}.  Thus,
  $B,I(\pi) \models \eq{\tau}$ for every $\tau\in\BB(\sigma)$. By \lemref{lem-Iequivalence},
  $I(\pi)$ contains $J(\sigma)$ as a conjunct; therefore, $B,I(\pi) \models \eq{\sigma}$. Since
  $\sigma$ here is an arbitrary element of $\AA(\pi)$, the second claim of
  \lemref{lem-premises} finishes the proof. \qed

\subsubsection*{Proof of \thmref{thm-correctness}}

The algorithm terminates because all pertinent functions have been proven
terminating.
  
Let $s\neq t$ be the disequality obtained in the step \ione\ of the
algorithm. Let $\pi$ be the path $\path st$, and let $\path st = \pi_1\theta\pi_2$, as
in the definition of $I'(\pi)$.
The two cases to consider, $s\neq t\in B$ and $s\neq t\in A$, will be referred to as Cases
1 and 2 respectively. Let $\varphi$ be the returned formula---$I(\pi)$ in Case 1;
$I'(\pi)$ in Case 2. 

\emph{(i) $\varphi$ is an $AB$-colorable conjunction of Horn clauses.}  
For any factor $\sigma$ of $\pi$
with $AB$-colorable endpoints, $J(\sigma)$ is an $AB$-colorable Horn
clause.  If $\pi$ has $B$-colorable endpoints, then so do all paths in $\PP(\pi)$
and so, by \lemref{lem-monotonicity}(iii), all paths in $\AA(\PP(\pi))$ have
$AB$-colorable endpoints. With \lemref{lem-Iequivalence}, this proves Case 1.
For Case 2, observe that if $\theta$ is empty, then $I(\theta)=\eq{\theta}=\true$; otherwise,
$\theta$ has $AB$-colorable endpoints. Also, $\pi_1$ and $\pi_2$ are $A$-paths, so by
the dual of \lemref{lem-monotonicity}(iii), all paths in $\BB(\pi_1)\cup\BB(\pi_2)$
have $AB$-colorable endpoints. These facts suffice to derive the proof of Case 2
from the already proved Case 1.

\emph{(ii)  $A\models\varphi$.} 
By the first claim of \lemref{lem-premises}, $A \models J(\sigma)$ holds for
every path $\sigma$. This suffices for Case 1. For Case 2 then, we only need to check
that the last conjunct of $I'(\pi)$ is implied by $A$, which amounts to showing
$A,\eq{\BB(\pi_1)},\eq{\BB(\pi_2)} \models \lnot\eq{\theta}$. This indeed follows from the first
claim of \lemref{lem-premises}, the transitivity entailment $\eq{\pi_1}, \eq{\theta},
\eq{\pi_2} \models \eq{\pi}$, and the assumption $\lnot\eq{\pi}\in A$.

\emph{(iii) $B,\varphi \models \false$.} 
In Case 1, we have $\lnot\eq{\pi}\in B$, so
\lemref{lem-main} finishes the proof. In Case 2, \lemref{lem-main} implies
$B,I'(\pi) \models \eq{\theta}$ and $B,I'(\pi) \models I(\tau)$ for every
$\tau\in\BB(\pi_1)\cup\BB(\pi_2)$. These consequences of $B\cup I'(\pi)$ contradict the
last conjunct of $I'(\pi)$. 
\qed


\section{Comparison with McMillan's Algorithm}
\label{sec-comparison}

\begin{figure}
  \centering
\includegraphics[scale=0.6]{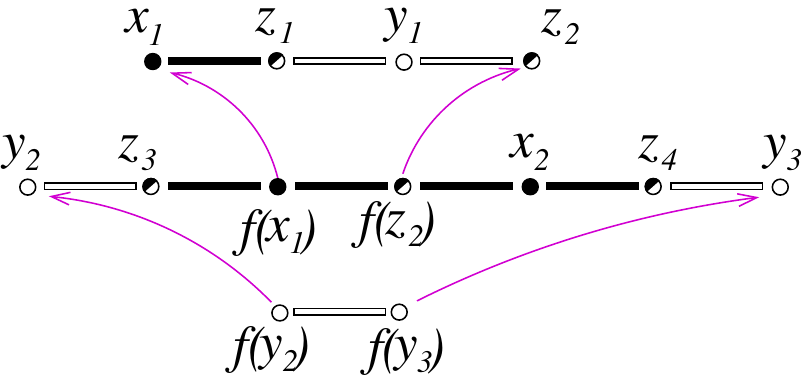}      
  \caption {
A colored congruence graph for $ A=\{x_1=z_1, z_3=f(x_1), f(z_2)=x_2, x_2=z_4\}$
and $B=\{z_1=y_1, y_1=z_2, y_2=z_3, z_4=y_3, f(y_2)\neq f(y_3)\}$ with two derived
edges $\langle f(x_1),f(z_2)\rangle$ and $\langle f(y_2),f(y_3)\rangle$.  }
  \label{fig-newvertex}
\end{figure}

\def\mcm{\cite{McM-TCS-05}}

Our \EUF ground interpolation algorithm is, as far as we know, the only
alternative to McMillan's algorithm \mcm. The latter constructs an interpolant
for $A,B$ from the proof of $A,B\models\false$ derived in a formal system ($\EE$,
say) with rules for introducing hypotheses (equalities from $A\cup B$),
reflexivity, symmetry, transitivity, congruence, and contradiction (deriving
$\false$ from an equality and its negation). The algorithm proceeds top down by
annotating each intermediate derived equality $u=v$ (or $\false$ in the final
step) with a quadruple of the form $[u',v',\rho,\gamma]$, where $u',v'$ are terms and
$\rho,\gamma$ are $AB$-colorable formulas. The annotation of each derived equality is
obtained from annotations of the equalities occurring in the premises of the
corresponding rule application. The exact computation of annotations is
specified by 11 rules, each corresponding to a case (depending on colors of the
terms involved) of one of the original six rules. An invariant that relates a
derived intermediate equality with its annotation is formulated and all 11 rules
are proved to preserve the invariant.  The invariant implies that if
$[u',v',\rho,\gamma]$ is the annotation of $\false$, then $\rho\Rightarrow\gamma$ is an interpolant
for $A,B$. It can be shown that $\rho$ is always a conjunction of Horn clauses,
and $\gamma$ is a conjunction of equalities and at most one disequality.

There is a clear relationship between proofs in the formal system $\EE$ and
congruence graphs from which our interpolants are derived. The main difference
is that in congruence graphs, paths condense inferences by reflexivity,
symmetry, and transitivity. A congruence graph provides a big-step proof that,
if necessary, can be expanded into a proof in the system $\EE$.

In \exref{ex-transitivity} (\figref{fig-transitivity}(a)) our algorithm looks at
the path $\path{y_3}{z_4}$, summarizes its only $A$-factor, producing the
interpolant $z_1 = z_4$. McMillan's algorithm processes the path edge-by-edge,
eagerly summarizing $A$-chains with $AB$-colorable endpoints, so that the
interpolant it produces is $z_1=z_2 \land z_2=f(z_3) \land f(z_3)=z_4$.

For the second difference, consider \exref{ex-mainex} (\figref{fig-mainex}(b))
where McMillan's algorithm produces an entangled version $(z_1=z_2 \land (z_3=z_4
\Rightarrow z_5=z_6))\Rightarrow z_3=z_4 \land z_7=z_8$ of our interpolant $ (z_1=z_2\Rightarrow z_3=z_4) \land
(z_5=z_6\Rightarrow z_7=z_8)$, computed in \exref{ex-run}. In general, McMillan's
algorithm accumulates $B$-justifications (duals of our $J(\sigma)$) in the $\rho$-part
of the annotation and keeps them past their one-time use to derive a particular
conjunct of $\gamma$.

The third difference is in creating auxiliary $AB$-terms (``equality
interpolants'', in the terminology of Yorsh and Musuvathi~\cite{YorMus-CADE-05}) to split
derivations of equalities in which one side is not $A$-colorable and the other
is not $B$-colorable, as in \exref{ex-uncolorable}. 
We introduce such terms in the preliminary step (\textbf{i2}) of our algorithm
only when required to make the congruence graph colorable.
In contrast,
McMillan's algorithm introduces these terms ``on-the-fly'', as in the example
illustrated in \figref{fig-newvertex}. When it derives the equality $x_1=z_2$,
its annotation is $[z_1,z_2,\true,\true]$, then when it uses the congruence rule
to derive $f(x_1)=f(z_2)$, this equality gets annotated with
$[f(z_1),f(z_2),\true,\true]$, and the term $f(z_1)$ becomes part of the final
interpolant $z_3=f(z_1)\land f(z_2)=z_4$. On the other hand, our algorithm
recognizes the edge $\langle f(x_1),f(z_2)\rangle$ as $A$-colorable and does not split it;
the interpolant it produces is $z_1=z_2\Rightarrow z_3=z_4$.

The final difference is in flexibility. McMillan's algorithm is fully specified
and leaves little room for variation. On the other hand, the actions in the step
(\textbf{i2}) of our algorithm are largely non-deterministic.  Our current
implementation chooses to minimize the number of vertices in the colorable
modification of $\Gamma$, and then colors the graph with a strategy that eagerly
minimizes the number of factors in the relevant paths. Other choices are yet to
be explored.

\subsection{Experimental evaluation}

\begin{figure}
  \centering
 
  \begin{minipage}{0.495\linewidth}
   { \includegraphics[angle=0,width=\linewidth]
     {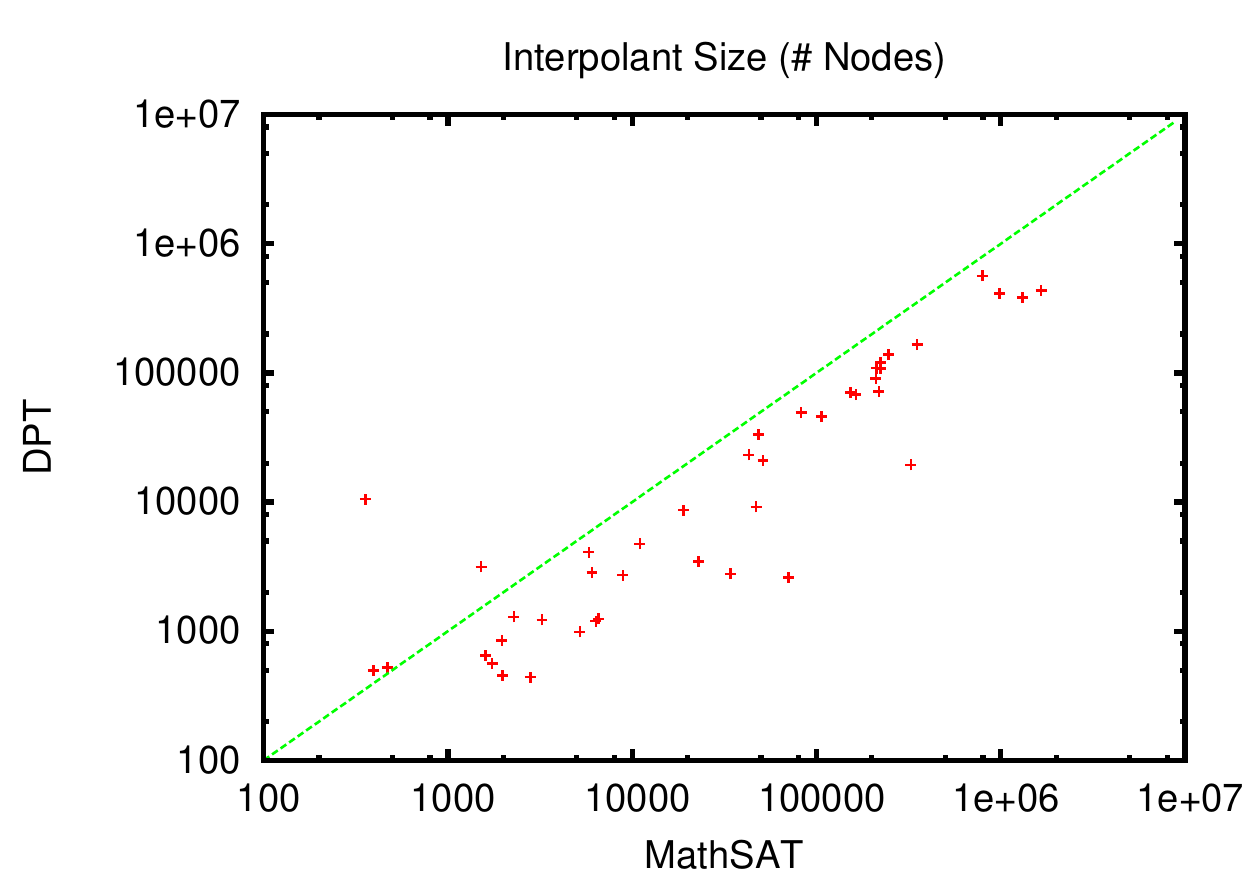}}
 \end{minipage}
 \begin{minipage}{0.495\linewidth}
   { \includegraphics[angle=-90,width=\linewidth]
     {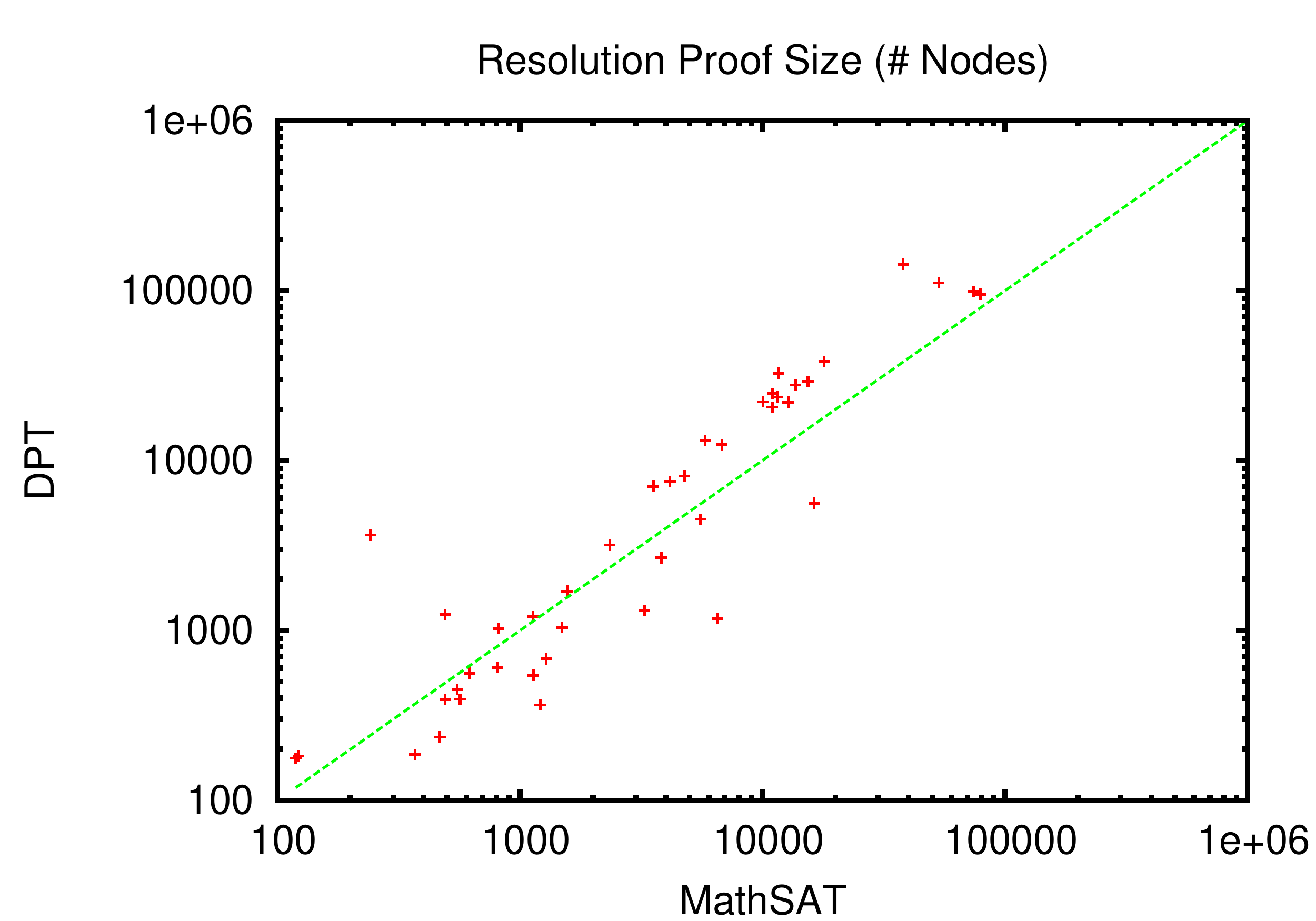}}  
 \end{minipage}

  \caption{\emph{DPT vs.~MathSAT} on  45 benchmarks from the
    \emph{MathSAT} library derived by partitioning unsatisfiable SMT-LIB
    benchmarks \cite{BarST-SMTLIB}.}
  \label{fig-comparison}
\end{figure}

In general, our interpolation algorithm produces smaller and simpler interpolants. For
experimental confirmation, we used the state-of-the-art implementation of
McMillan's algorithm in \emph{MathSAT} \cite{CimGS-TACAS-08} and compared it
against our interpolation-generating extension of the \emph{DPT} solver~\cite{DPT}.  

Two other relevant components---the propositional interpolation
algorithm, and the algorithm for combining propositional and theory
interpolation in a $\textsc{DPLL}(\TT)$ framework
\cite{McM-TCS-05,CimGS-TACAS-08}---are the same in \emph{MathSAT} and
\emph{DPT}, and therefore unlikely to substantially affect the comparison.  The
last factor to be accounted for in this comparison is the size of the resolution
proofs derived from the $\textsc{DPLL}$ search within each solver. 
Since these sizes are comparable, 
we can eliminate differences in propositional reasoning as
a cause for \emph{DPT}'s producing smaller interpolants.

We ran both solvers on 45 \EUF interpolation benchmarks selected from the set of
100 that are used in \cite{CimGS-TACAS-08}. (In the remaining 55 benchmarks,
either all formulas in $A$ are $B$-colorable, or all formulas in $B$ are
$A$-colorable, so one of the formulas $A$, $\lnot B$ is an easily obtained
interpolant.)  Both solvers computed 42 interpolants, timing out in 100\emph{s}
on the same three benchmarks.  Runtimes were comparable, with \emph{DPT} being
slightly faster.  \figref{fig-comparison} shows the sizes of interpolants
produced: \emph{DPT} interpolants are, on average, 3.8 times smaller, in spite
of \emph{DPT} proofs being, on average, 1.7 times larger.

While these experimental results confirm 
the claim that 
our algorithm produces smaller interpolants,
we observe that 
formula size is not necessarily a good metric,
given the ability of modern SMT-solvers to process large formulas quickly.
It could be argued that 
some measure of \emph{logical strength} would be better instead. 
The case for that, however, is not obvious either.
To start, the only reasonable way to compare two first-order logic formulas 
$\varphi_1$ and $\varphi_2$
for logical strength is to check whether 
one of the two entails the other in the theory
(\ie, whether $\varphi_1 \models_\TT \varphi_2$ or
$\varphi_2 \models_\TT \varphi_1$).
Unfortunately, entailment is not a total relation
and so it is possible to have incomparable interpolants 
for the same partition $A,B$ of a set of formulas.
Finally, even with comparable invariants,
whether the stronger or the weaker one is better
depends on the specific application using them;
worse still, for other applications, 
such as interpolation-based predicate abstraction,
it is arguable that logical strength (or formula size for that matter)
is of any importance,
since interpolants are simply mined for useful predicates.
Further work is needed to identify useful evaluation metrics for interpolants
and then see if the flexibility of our algorithm,
or a suitably modified version of it,
can be used to produce better interpolants according to some of those metrics.


\section{Interpolation as a Cooperative Game}
\label{sec-game}

Our results about \EUF interpolation
can be generalized to a wider class of theories $\TT$
in terms of a cooperative \emph{interpolation game}
between two deductive provers 
for \TT---possibly two copies of the same prover.
The game metaphor suggests a simple and general mechanism 
for producing interpolants
from sets of formulas and theories that satisfy certain requirements.
We define this mechanism and prove its properties in \secref{sec-int-runs} and \secref{sec-runs},
after giving an informal general description of the interpolation game.

For the rest of the section,
let $\TT$ be a first-order theory of signature $\Sigma$, and
let $A$ and $B$ be two disjoint sets of formulas possibly containing 
\defemph{free symbols}, \ie,
predicate and function symbols not in $\Sigma$.  
For convenience, and without loss of generality,
we consider only formulas with no free variables.
Let $\Sigma_I$ be the \defemph{shared signature}, 
the expansion of $\Sigma$ with the free symbols
occurring in both $A$ and $B$.

\subsection{The interpolation game}

The participants are an \defemph{$A$-prover} and a \defemph{$B$-prover}
which incrementally construct 
a set $S_A$ and a set $S_B$ of $\Sigma_I$-formulas. 
The game starts with $S_A = S_B = \emptyset$ and 
proceeds in rounds so that at each round one of the following happens:
\begin{iteMize}{$\bullet$}
\item
the $A$-prover adds to $S_A$ one or more $\Sigma_I$-formulas $\alpha$ 
such that $A, \beta_1, \ldots, \beta_n \tentails \alpha $
for some $\beta_1, \ldots, \beta_m \in S_B$,
the \defemph{$B$-premises} of $\alpha$;
\item 
the $B$-prover adds to $S_B$ one or more $\Sigma_I$-formulas $\beta$
such that $B, \alpha_1, \ldots, \alpha_n \tentails \beta$
for some $\alpha_1, \ldots, \alpha_n \in S_A$,
the \defemph{$A$-premises} of $\beta$.
\end{iteMize}

\noindent
The game ends successfully 
when the $B$-prover adds $\false$ to $S_B$.  

As we discuss below,
a $\TT$-interpolant for $A$ and $B$ can be generated 
from a successful run of the game by tracking 
the $B$-premises of each formula in $S_A$ and 
the $A$-premises of each formula in $S_B$. 

Note that, as described, the interpolation game involves 
arbitrary theories and input sets $A$ and $B$.
Also, the game does not have to use two provers literally.
If $A \cup B$ has a \emph{local} refutation (see later)
in the theory $\TT$,
it is possible to extract from that refutation
a successful run of the game
from which a $\TT$-interpolant of $A$ and $B$ can then be generated.

For some theories and classes of input formulas 
the game admits \emph{complete strategies}, guaranteed to end the game 
when $A$ and $B$ are jointly $\TT$-unsatisfiable.  
Depending on the theory and the class of input formulas, 
these strategies can be considerably restrictive 
in the choice of formulas to propagate from one prover to the other
(\ie, formulas to add to $S_A$ and $S_B$).
For instance, 
when $A$ and $B$ are sets of ground literals and 
the theory is \emph{convex},\footnote{
A theory is \defemph{convex} 
if $L\tentails p_1\lor\cdots\lor p_k$ implies $L\tentails p_i$ for some $i$, where $L$ is any set of ground literals and the $p_i$ are any positive literals.
} 
it is enough to propagate just ground atomic formulas 
in all rounds of the game. 
In that case, 
all interpolants computed will be conjunctions of ground Horn clauses.
 
The interpolation method described in \secref{sec-interpolation} 
for the theory of equality (which is convex) can be seen 
as a customized implementation of the interpolation game, 
with formula propagation restricted to (positive) equalities.  
A colorable congruence graph is a compact representation of a local refutation,
and the interpolation function $I$ defined in \secref{sec-interpolation}
can be understood as generating the interpolant 
from a successful run of the game extracted from the local refutation.

\begin{figure}
\centering
\includegraphics[scale=0.6]{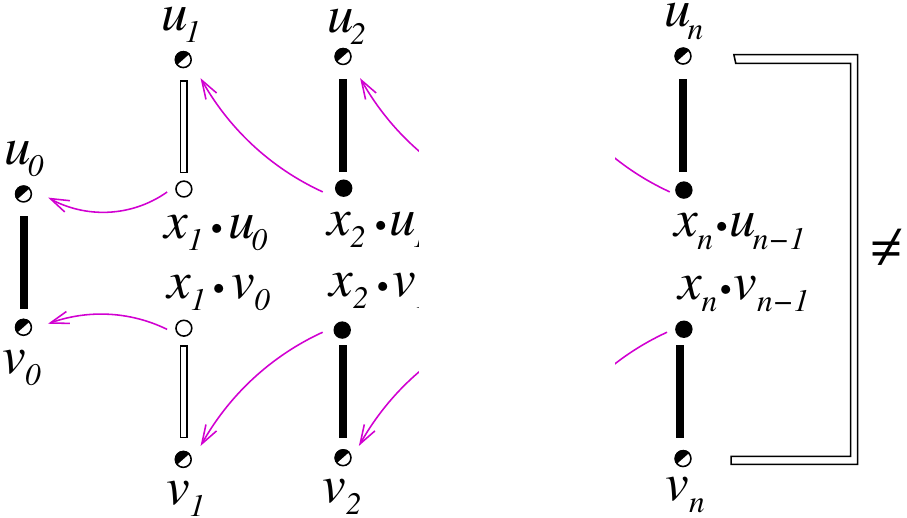}    
\caption{A long derivation. (\exref{ex-yo})}
\label{fig-long}
\end{figure}

\begin{exa}
\label{ex-transitivity-rev}
Looking back at Example~\ref{ex-transitivity}
in terms of the interpolation game above, 
we can see that in each of the three cases presented in the example
there is a successful interpolation game with two rounds.
In the first round, 
the $A$-prover derives a conjunction of literals in the shared signature
(respectively, $z_1=z_4$,\,
$z_1=z_2 \land f(z_3)=z_4 \land f(z_2)=z_3$
and 
$z_1=z_2 \land f(z_3)\neq z_4 \land f(z_2)=z_3$)
that the $B$-prover uses them to derive $\false$.
\qed
\end{exa}

\begin{exa}
\label{ex-first-horn-rev}
For the sets $A$ and $B$ in Example~\ref{ex-first-horn},
there is a game with three rounds where 
$u_0 = v_0$ is initially derived by the $B$-prover, 
$u_1=v_1$ is derived next by the $A$-prover, 
and then $\false$ is derived by the $B$-prover.
\qed
\end{exa}

\begin{exa}
\label{ex-yo}
Generalizing the previous example, consider this matrix---organized set of
literals:
 $$\xymatrix@C=.42cm
{u_0 = v_0 & 
{
  \begin{array}{c}
 x_1\cdot u_0 = u_1 \\  x_1\cdot v_0 = v_1
  \end{array}
}
 &
{ \begin{array}{c}
x_2\cdot u_1 = u_2 \\
x_2\cdot v_1 = v_2
  \end{array}
}
&
\ldots
&
{  \begin{array}{c}
x_n\cdot u_{n-1} = u_n \\
x_n\cdot v_{n-1} = v_n
  \end{array}
}
& u_n \neq v_n
}
$$
Let $A$ be the set of equalities occurring in the odd-numbered columns 
(with columns counted starting from 1) of this matrix, 
and $B$ be the set of the remaining
equalities; see \figref{fig-long}. 
The shared symbols are $u_0,v_0,\ldots,u_n,v_n$,
the symbols local to $A$ are $x_2,x_4,\ldots$, and the symbols local to $B$ are
$x_1,x_3,\ldots$

A run of the interpolation game takes $n+2$ rounds. 
It begins with the $A$-prover adding $u_0=v_0$ to
$S_A$. Then, using the equalities from the second column, the $B$-prover can
derive $u_1=v_1$, and add it to $S_B$. Now, the $A$-prover can use this equality
together with equalities from the third column to derive $u_2=v_2$ and add it to
$S_A$.  Assuming $n$ is even, the last equality $u_n=v_n$ will be derived by the
$A$-prover, after which $B$ derives $\false$. Collecting justifications of all
equalities derived by $A$, we obtain the interpolant
\begin{equation*}
  (u_0=v_0) \land (u_1=v_1\limplies u_2=v_2) \land \cdots \land
(u_{n-1}=v_{n-1} \limplies u_{n}=v_{n})\ .
\end{equation*}
\qed
\end{exa}

\begin{rem}
The well-known method for combining decision
procedures due to Nelson and Oppen~\cite{NelOpp-TPLS-79} 
is essentially a version of the interpolation game.  
The main differences are that in the Nelson-Oppen method
   (i) the input
  sets of formulas $A_1$ and $A_2$ need not be jointly $\TT$-unsatisfiable; (ii)
  the goal is not to produce interpolants for $A_1$ and $A_2$ but just to
  check the \TT-unsatisfiability of $A_1 \cup A_2$; (iii) \TT is the union of
  two signature-disjoint theories $\TT_1$ and $\TT_2$; (iv) each formula $A_i$
  is built from the symbols of $\TT_i$ and free constants; (v) each $A_i$-prover
  works just over $\TT_i$ instead of the whole \TT; (vi) additional restrictions
  on $\TT_1$ and $\TT_2$ guarantee termination even when $A_1 \cup A_2$ is
  \TT-satisfiable.  
  
  A description of a Nelson-Oppen combination framework in
  terms similar to our interpolation game 
  is given by Ghilardi~\cite{Ghi-JAR-05}.
\qed
\end{rem}

\subsection{Extracting interpolants from interpolation runs}
\label{sec-int-runs}

To show how to generate $\TT$-interpolants from runs of the interpolation game
we start by formalizing the notion of a run.

\begin{defi}
\label{def-run}
A \defemph{$\TT$-interpolation run} for $A$ and $B$
is a triple $(S_A, S_B, \sqsubset)$ 
where
$S_A$ and $S_B$ are two disjoint finite sets of $\Sigma_I$-formulas 
and 
$\sqsubset$ is a well-founded (partial) ordering on $S_A \cup S_B$
with associated computable functions
$\pr{B}:S_A \to 2^{S_B}$, $\pr{A}:S_B \to 2^{S_A}$ such that:

\begin{enumerate}[(1)]
\item
$A, \pr{B}(\alpha) \tentails \alpha$
\ and \ 
$\beta \sqsubset \alpha$ for all $\beta \in  \pr{B}(\alpha)$;

\item
$B, \pr{A}(\beta) \tentails \beta$
\ and \ 
$\alpha \sqsubset \beta$ for all $\alpha \in  \pr{A}(\beta)$.
\end{enumerate}
A $\TT$-interpolation run $(S_A, S_B, \sqsubset)$ is \defemph{successful}
if $\false \in S_B$.

\end{defi}
\bigskip

Given a $\TT$-interpolation run $(S_A, S_B, \sqsubset)$,
we extend $\pr{B}$ from $S_A$ to $2^{S_A}$ as done in~\secref{sec-proof},
that is,
for all $S \subseteq S_A$, 
\[
\pr{B}(S) = \bigcup \{\pr{B}(\alpha) \mid \alpha \in S\} \ .
\] 
We extend $\pr{A}$ from $S_B$ to $2^{S_B}$ in a similar way.
Then, for all $\beta \in S_B$ let
\[
\pr{}(\beta) = \{\beta\} \cup \bigcup\{\pr{}(\beta') \mid \beta' \in \pr{B}(\pr{A}(\beta))\} \ .
\]
Extending $\pr{}$ to $2^{S_B}$ as done with $\pr{A}$,
we can write the definition of $\pr{}$ more compactly as
\[
\pr{}(\beta) = \{\beta\} \cup \pr{}(\pr{B}(\pr{A}(\beta))) \ .
\]

Finally, we define the (computable) function $\II$ 
from $S_B$ to the set of $\Sigma_I$-formulas such that
\[
\II(\beta) = \bigcup\{ \pr{B}(\alpha) \limplies \alpha \mid \alpha \in \pr{A}(\pr{}(\beta))\} \ .
\]

\noindent 
This function returns \defemph{partial $\TT$-interpolants} in the following
sense.

\begin{lem}
\label{lem:gen-int}
Let $(S_A, S_B, \sqsubset)$ be a $\TT$-interpolation run for $A$ and $B$
and 
let $\II$ be defined as above. 
Then, for all $\beta \in S_B$,
\begin{enumerate}[\em(1)]
\item
$A \tentails \II(\beta)$;
\item
$B, \II(\beta) \tentails \beta$.
\end{enumerate}

\end{lem}
\proof
We prove both claims by well founded induction on $\sqsubset$.
By definition,
$\pr{}(\beta) = \{\beta\} \cup \pr{}(\beta_1) \cup \cdots \cup \pr{}(\beta_k)$
where $\{\beta_1, \ldots, \beta_k\} = \pr{B}(\pr{A}(\beta))$ with $k \geq 0$.
Then, 
\[
\begin{array}{lll}
 \II(\beta) 
 & = & \bigcup\{ \pr{B}(\alpha) \limplies \alpha \mid \alpha \in \pr{A}(\{\beta\} \cup \pr{}(\beta_1) \cup \cdots \cup \pr{}(\beta_k))\} 
 \\[1ex]
 & = & \bigcup\{ \pr{B}(\alpha) \limplies \alpha \mid \alpha \in \pr{A}(\beta) \cup \pr{A}(\pr{}(\beta_1)) \cup \cdots \cup \pr{A}(\pr{}(\beta_k))\} 
 \\[1ex]
 & = & \bigcup\{ \pr{B}(\alpha) \limplies \alpha \mid \alpha \in \pr{A}(\beta)\} \cup
       \bigcup_i\bigcup\{ \pr{B}(\alpha) \limplies \alpha \mid \alpha \in \pr{A}(\pr{}(\beta_i))\} 
 \\[1ex]
 & = & \bigcup\{ \pr{B}(\alpha) \limplies \alpha \mid \alpha \in \pr{A}(\beta)\} \cup
       \bigcup_i\II(\beta_i) 
 \\[1ex]
\end{array}
\]

\changed{
To prove Claim (1), we check that every element of $\II(\beta)$ is entailed by $A$.
Indeed, $A \tentails \II(\beta_i)$ holds by the induction hypothesis, and $A
\tentails \pr{B}(\alpha) \limplies \alpha$ follows directly from the defining
property of $\pr{B}$.
}

The defining property $B, \pr{A}(\beta) \tentails \beta$ of $\pr{A}$ reduces
proving Claim (2) to proving $B, \II(\beta) \tentails \alpha$, for every
$\alpha\in\pr{A}(\beta)$. Since $(\pr{B}(\alpha) \limplies \alpha)$ is in $\II(\beta)$,
it suffices to prove that $B, \II(\beta) \tentails \pr{B}(\alpha)$. And indeed,
$\pr{B}(\alpha)$ is a subset of $\{\beta_1, \ldots, \beta_k\}$, and $B,
\II(\beta) \tentails \beta_i$ holds by induction hypothesis.
\qed
\bigskip

Lemma~\ref{lem:gen-int} is the induction vehicle for the following main result.

\begin{thm}
  Let $(S_A, S_B, \sqsubset)$ be a successful $\TT$-interpolation run for $A$
  and $B$ and let $\II$ be defined as above.  The formula $\bigwedge
  \II(\false)$ is a $\TT$-interpolant of $A$ and $B$.
\end{thm}
\proof Since $\false\in S_B$, we can instantiate Lemma~\ref{lem:gen-int} with
$\beta$ equal to $\false$. The free symbols occurring in $\II(\false)$ are shared by $A$ and
$B$ because, by construction, $\II$ returns $\Sigma_I$-formulas.  \qed

\subsection{Interpolation runs from local refutations}
\label{sec-runs}

The next question is how to construct successful interpolation runs 
for $A$ and $B$.
One way is to extract them from proofs of $\TT$-unsatisfiability 
of $A \cup B$ in a suitable proof system.
We define a fairly general notion of a proof system 
and show that any refutation of $A \cup B$ in the system
that is local in the sense of Jhala and McMillan~\cite{JhaMcM-TACAS-06} 
contains a successful interpolation run.

\subsubsection{Proofs and proof systems}
A \defemph{proof rule} is a binary relation between finite sets of formulas and formulas.
Any pair $\varphi_1, \ldots, \varphi_n \vdash \varphi$, with $n\geq 0$,
in a proof rule, usually written as
\[
\begin{array}{c}
 \varphi_1 \quad \cdots \quad \varphi_n \\
 \hline
 \varphi
\end{array} \ ,
\]
is an 
\changed{\defemph{inference step}}
with \defemph{premises} $\varphi_1, \ldots, \varphi_n$ and \defemph{conclusion} $\varphi$.
The conclusion of an inference step with an empty set of premises is an \defemph{axiom}.
A \defemph{proof system} is a set of proof rules.
A proof rule is \defemph{sound} with respect to a theory $\TT$
if $\varphi_1, \ldots, \varphi_n \tentails \varphi$
for each inference step $\varphi_1, \ldots, \varphi_n \vdash \varphi$ of the rule.

\begin{defi}
For every proof system $\RR$, formula $\varphi$ and set of formulas $S$,
a \defemph{proof of $\varphi$ from $S$ in $\RR$} is a labelled tree 
defined inductively as follows.

\begin{enumerate}[(1)]
\item
If $\varphi \in S$,
the one-node tree with root labelled $\varphi$ is a proof 
of $\varphi$ from $S$ in $\RR$;

\item
if $\varphi_1, \ldots, \varphi_n \vdash_\RR \varphi$ is 
an inference step of $\RR$ and 
$\D_i$ a proof of $\varphi_i$ from $S$ in $\RR$ for $i=1,\ldots,n$,
then the tree $\D$ with root $\varphi$ and 
immediate subtrees $\D_1, \ldots, \D_n$ is 
a proof of $\varphi$ from $S$ in $\RR$.
The roots of $\D_1, \ldots, \D_n$ are the \defemph{parents} of the root of $\D$.
\end{enumerate}

A \defemph{refutation} of $S$ in $\RR$ is a proof of $\false$ 
from $S$ in $\RR$.
\qed
\end{defi}

In the following,
we will identify nodes of a proof with their labels 
when this does not cause confusion.
\changed{
Observe that if all the rules of $\RR$ are sound 
with respect to a theory $\TT$,}
then
$S \tentails \varphi$ for each proof in $\RR$ 
of a formula $\varphi$ from a set of formulas $S$.
In particular, 
any set of formulas that has a refutation in $\RR$ is $\TT$-unsatisfiable.

Extending the terminology introduced in \secref{sec-ccg}, 
we say that 
an inference step in $\RR$ is \defemph{$A$-colorable} 
(resp., \defemph{$B$-colorable}) 
if the formulas in the inference step are all $A$-colorable 
(resp., all $B$-colorable).
We define a proof of a formula $\varphi$ from $A \cup B$ in
$\RR$ to be \defemph{local} 
if every inference step in the proof is $A$- or $B$-colorable.
An example of local proof is shown in Figure~\ref{fig-ref}.

\subsubsection{Constructing interpolation runs}
\label{sec-cons-int-runs}

Fix any proof system $\RR$ that is sound for $\TT$.
We show that from any local proof $\D$ from $A \cup B$ in $\RR$,
we can construct a $\TT$-interpolation run $(S_A, S_B, \sqsubset)$ 
so that if $\D$ is a refutation then $\false \in S_B$.
(Then, 
the function $\II$ can be used to produce 
a $\TT$-interpolant of $A$ and $B$
as shown in \secref{sec-int-runs}.)

Let $\D$ be a local refutation of $A \cup B$ in $\RR$.
Without loss of generality we can assume that
$(i)$
if two nodes of $\D$ have the same label,
then they are roots of structurally identical subtrees of $\D$,
and
$(ii)$
the parents of $\false$ in $\D$ are all $B$-colorable. 
Local refutations that do not satisfy Requirement $(ii)$
can be modified
by replacing $\false$ with a new logical constant $\false'$ 
interpreted in the same way and 
then adding a final, $B$-colorable inference, $\false' \vdash \false$. 

Define $\sqsubset$ as the relation on the labels of $\D$ such that 
$\varphi \sqsubset \psi$ iff $\varphi$ is an ancestor of $\psi$ in $\D$.
By the assumptions on $\D$,
the (finite) relation $\sqsubset$ is acyclic. 
Hence, both $\sqsubset$ and its inverse are well founded.

If we cut $\D$ at a node $\varphi$, 
we obtain two local proofs in $\RR$:
a local proof of $\varphi$ from $A\cup B$ 
(the tree rooted at $\varphi$),
and a local proof of $\false$ from $A\cup B \cup \{\varphi\}$
(the remaining tree, with same root as $\D$ and $\varphi$ as one of its leafs).
More generally, 
we can decompose $\D$ into several smaller local proofs by cutting 
it repeatedly at different nodes. 

\begin{defi}
A pair $T_A, T_B$ of sets of nodes in $\D$ is a \defemph{coloring cut} of $\D$ 
if
\begin{enumerate}[(1)]
\item  
all nodes in $T_A\cup T_B$ are $AB$-colorable;
\item 
$T_A$ and $T_B$ are disjoint, and $\false$ is in $T_B$;
\item 
for all $\alpha\in T_A$ and $\psi\in T_A\cup (B \setminus T_B)$ 
with $\psi\sqsubset\alpha$, 
there is a $\beta\in T_B$ such that $\psi\sqsubset\beta\sqsubset\alpha$;
\item 
for all $\beta\in T_B$ and $\psi\in T_B\cup (A \setminus T_A)$ 
with $\psi\sqsubset\beta$, 
there is a $\alpha\in T_A$ such that $\psi\sqsubset\alpha\sqsubset\beta$.
\end{enumerate}
\end{defi}

It is simple to verify 
that cutting $\D$ at the nodes of $T_A\cup T_B$, 
where $T_A,T_B$ is a coloring cut, decomposes $\D$ into colorable proofs. 
More precisely, 
every resulting smaller proof rooted at a node of $T_A$ (resp., $T_B$) 
consists of $A$-colorable (resp., $B$-colorable) nodes.

\begin{figure}
\small
\newcommand{\Paaab}{
 \infer{\ac{q(f a, a)}}
       {\ac{r(b) \lor q(f a, a)} & \boxed{\lnot r(b)} }
}
\newcommand{\Paaa}{
 \infer{\ac{p(a) \limplies r(a)}}
       {\ac{\bar{\forall}(p(u) \land q(v,u) \limplies r(u))} & \Paaab}
}
\newcommand{\Paabb}{
 \infer{\bc{\bar{\forall}(s(v) \land r(v) \limplies t(f v))}}
       {\bc{\bar{\forall}(s(v) \land r(v) \limplies r(f v))}}
}
\newcommand{\Paab}{
 \infer{\boxed{\bar{\forall}(r(x) \limplies t(f x))}}
       {\bc{\bar{\forall}s(x)} & \Paabb}
}
\newcommand{\Paa}{
 \infer{\ac{p(a) \limplies t(f a)}}
       {\Paaa & \Paab}
}
\newcommand{\Pa}{
 \infer{\boxed{t(f a)}}
       {\Paa & \ac{p(a)}}
}

\[
 \infer{\boxed{\false}}{
 \Pa
 &
 \lnot t(f a)
}
\]
\caption{A local refutation $\D$ of $A \cup B$.
The symbols $r$ and $t$ are from the theory's signature.
Of the remaining symbols, $p$ and $q$ occur only in $\ac A$,
$s$ occurs only in $\bc B$, and $a, b$ and $f$ occur in both.
The boxed formulas are those in the coloring cut in Example~\ref{ex-cut}.
}
\label{fig-ref}
\end{figure}

\begin{exa}
\label{ex-cut}
\changed{
Let $\TT$ be some arbitrary theory 
with a signature consisting of the predicate symbols $r, t$
and such that $\tentails \forall x.(r(x) \limplies t(x))$.
Then, let
\begin{eqnarray*}
A & = & \{\forall u.(p(u) \land q(v,u) \limplies r(u)),\; p(a),\; r(b) \lor q(f a, a)\},
\\
B & = & \{ \forall v.(s(v) \land r(v) \limplies r(f v)),\; \forall x.s(x),\; \lnot r(b),\; \lnot t(f a)\} \ .
\end{eqnarray*}
where $p, q, a, b, f$ and $s$ are non-theory symbols.
The proof in Figure~\ref{fig-ref} is a local refutation of $A \cup B$.
The exact proof system used to build the refutation is not important here.
Simply observe that each inference step is sound with respect to $\TT$,
which shows that $A \cup B$ is $\TT$-unsatisfiable.  
}

A coloring cut of $\D$ is given by the sets
\begin{eqnarray*}
T_A  = \{t(f a)\}
& \text{ and } &
T_B = \{\lnot r(b), \forall x.(r(x) \limplies t(f x)), \false \}\ .
\end{eqnarray*}
Note that the last inference step (the one with conclusion $\false$) is both
$A$- and $B$-colorable.
In the cut, however, it is essentially seen as a $B$-colored step. 
\qed
\end{exa}
\medskip

Every coloring cut induces a successful interpolation run.

\begin{thm}
If $S_A,S_B$ is a coloring cut, 
then $(S_A, S_B, \sqsubset)$ is a successful $\TT$-interpolation run for $A$ and $B$.
\end{thm}
\proof 
It is enough to define functions $\pr{A}$ and $\pr{B}$ satisfying 
Definition~\ref{def-run}.

For each $\alpha\in S_A$, 
let $\D_\alpha$ be the proof of $\alpha$ 
in the decomposition of $\D$ defined by the coloring cut $S_A,S_B$.
Define
\[
\pr{A}(\alpha) = \{\beta \in S_B \mid \beta \text{ is a leaf of } \D_\alpha\} \ .
\]
Clearly, $\beta \sqsubset \alpha$ for all $\beta \in \pr{A}(\alpha)$.
To show that $A, \pr{A}(\alpha) \tentails \alpha$
we show that every leaf of $\D_\alpha$ is in $A\cup \pr{A}(\alpha)$.
Now,
every leaf $\varphi$ of $\D_\alpha$ is 
either a leaf of $\D$ (so an element of $A\cup B$), 
or a cut node (an element of $S_A\cup S_B$). 
Since $\varphi\sqsubset\alpha$, it follows from the third defining
property of coloring cuts, that  $\varphi\notin S_A\cup B$
(otherwise, we would be able to cut $\D_\alpha$ at a node between 
$\varphi$ and $\alpha$). 
Thus, $\varphi\in S_B\cup A$. 

The function $\pr{B}$ is defined similarly.
\qed
\bigskip

\begin{exa}
\changed{
The $\TT$-interpolation run induced by the coloring cut in Example~\ref{ex-cut}
can be described informally in terms of the interpolation game as follows.
In the first round,
the $B$-prover adds to $S_B$
the formulas 
$\beta_1 = \lnot r(b)$ and $\beta_2 = \forall x.(r(x) \limplies t(f x))$,
each with an empty set of $A$-premises
(\ie, $\pr{A}(\beta_1) = \pr{A}(\beta_2) = \emptyset$).
In the second round,
the $A$-prover adds to $S_A$ the formula $\alpha_1 = t(f a)$,
with $\pr{B}(\alpha_1) = \{\beta_1, \beta_2\}$.
In the third and final round,
the $B$-prover adds $\false$ to $S_B$, with $\pr{A}(\false) = \{\alpha_1\}$.
}

The $\TT$-interpolant computed by the function $\II$,
defined in~\secref{sec-int-runs},
from this interpolation run is $\beta_1 \land \beta_2 \limplies \alpha_1$,
as shown below.
\[
\begin{array}{lll}
 \pr{}(\beta_1)
 & = & \{\alpha_1\} \cup \pr{}(\pr{B}(\pr{A}(\beta_1)))
   =   \{\alpha_1\} \cup \emptyset
  \\
 & = & \{\alpha_1\}
 \\[2ex]
 \pr{}(\beta_2))
 & = & \{\beta_2\} \cup \pr{}(\pr{B}(\pr{A}(\beta_2))))
   =   \{\beta_2\} \cup \emptyset
 \\
 & = & \{\beta_2\}
\\[2ex]
\pr{}(\false)
 & = & \{\false\} \cup \pr{}(\pr{B}(\pr{A}(\false)))
 \\
 & = & \{\false\} \cup \pr{}(\pr{B}(\alpha_1))
   =   \{\false\} \cup
       \pr{}(\beta_1) \cup
       \pr{}(\beta_2)
 \\
 & = & \{\false,\, \beta_1,\, \beta_2 \}
 \\[2ex]
\pr{A}(\pr{}(\false))
 & = & \pr{A}(\false) \cup \pr{A}(\beta_1) \cup \pr{A}(\beta_2)
   =  \{\alpha_1\} \cup \emptyset \cup \emptyset
 \\
 & = & \{\alpha_1\}
 \\[2ex]
\II(\false)
 & = & \bigcup\{ \pr{B}(\alpha) \limplies \alpha \mid \alpha \in \pr{A}(\pr{}(\false))\}
 \\
 & = & \{\beta_1 \land \beta_2 \limplies \alpha_1\}
\end{array}
\]

\qed
\end{exa}

We stress that computing a coloring cut from local refutations
is just one way to produce interpolation runs.
For specific theories, other mechanisms are possible.
A crucial point, however, is that 
local refutations always admit a coloring cut. 
In fact,
with $\D$ and with $\sqsubset$ as defined in~\secref{sec-cons-int-runs},
a coloring cut of $\D$ is provided by the sets $S_A$ and $S_B$ 
defined inductively as follows
over the set of $AB$-colorable nodes $\varphi$ of $\D$:
\begin{enumerate}[(1)]
\item
$\false \in S_B$;
\item
if $\varphi \sqsubset \beta$ for some $\beta \in S_B$,
$\varphi$ is a leaf from $A$ or has a non-$B$-colorable parent,
and $\varphi $ is $\sqsubset$-maximal with these properties\footnote{
That is,
there is no $AB$-colorable $\varphi'$ with a non-$B$-colorable parent
such that $\varphi \sqsubset \varphi' \sqsubset \beta$.
},
then $\varphi \in S_A$;
\item
if $\varphi \sqsubset \alpha$ for some $\alpha \in S_A$,
$\varphi$ is a leaf from $B$ or has a non-$A$-colorable parent,
and $\varphi $ is $\sqsubset$-maximal with these properties,
then $\varphi \in S_B$.
\end{enumerate}

Different coloring cut algorithms produce different interpolation runs, 
and therefore different interpolants. 
The inductive definition above aims at minimizing the cardinality of 
$S_A\cup S_B$\footnote{
In this sense, 
it is analogous to the congruence path factorization used in \secref{sec-coloredcg},
where each relevant path is broken into \emph{maximal} subpaths consisting 
of equally colored edges.
In both cases, the intent is to minimize the number of \emph{color switches},
so to speak---the number of factors in one case and the size of the coloring cut in the other.
}
and so is likely to produce smaller interpolants.
If there is a need to find interpolants optimal in some other sense, 
one can hope that 
the problem will translate into a meaningful optimization problem 
for coloring cuts. 


\section{Conclusion}

Our study of interpolation for the theory of equality was motivated by the
central role this theory plays in SMT solving, and by the practical
applicability of interpolant-producing SMT solvers
in model checking.  The algorithm we presented
is easy to implement on top of the standard congruence closure procedure. It
generates interpolants of a simple logical form and smaller size than those
produced by the alternative method.

We identified \emph{congruence graphs} as a convenient structure to represent
proofs in \EUF and to derive interpolants. The possibilities for global analysis
and transformations of these graphs go beyond what we have explored.  Our
algorithm provides a basis for further refinement and multiple implementations.
This flexibility may prove useful when the notion of interpolant quality is
better understood.

The heart of our algorithm---the generation of an interpolant from a suitably
colored congruence graph---is not \EUF-specific. We showed that behind it is a
general \emph{interpolation game} and a general mechanism for deriving
interpolants from suitably colored (\emph{local}) proofs.

\section*{Acknowledgement}
We thank Alberto Griggio for providing us with 
the interpolation benchmarks used in \cite{CimGS-TACAS-08}, 
and with a \emph{MathSAT} executable for benchmarking.
We also thank the anonymous reviewers for their thoughtful comments on
a preliminary version of this work, and
their suggestions for improving the presentation.

\bibliographystyle{alpha}
\bibliography{local}

\end{document}